\documentclass[10pt,twocolumn,twoside]{IEEEtran}
\usepackage[belowskip=-10pt,aboveskip=0pt]{caption}

\usepackage{subcaption}

\usepackage[normalem]{ulem}
\renewcommand{\sout}[1]{}

\ifCLASSINFOpdf
\else
\fi

\usepackage{float}
\usepackage[labelsep=period]{caption}

\usepackage{float}

\usepackage{mathtools, cuted}

\usepackage{tabularx}
\usepackage{amsmath}
\usepackage{amssymb}
\usepackage{mathalfa}
\usepackage{hyperref}

\usepackage{nccmath}

\newcommand{\defint}[4]{\int_{#1}^{#2}{#3} d{#4}}

\newcommand{\w}{\omega}

\newcommand{\B}{\beta}

\newcommand{\expn}[1]{e^{#1}}

\newcommand{\Ap}{A_p}
\newcommand{\bp}{b_p}
\newcommand{\Bu}{B_u}

\newcommand{\pconj}{\overline{p}}

\renewcommand{\xi}{x_i}

\newcommand{\level}[1]{\text{mag}\{{#1}\}}

\newcommand{\phase}[1]{\text{phase}\{{#1}\}}

\newcommand{\kB}{k_{\B}}

\renewcommand{\Re}{\text{Re}}
\renewcommand{\Im}{\text{Im}}

\newcommand{\sLaplace}{\mathsf{s}}
\newcommand{\pLaplace}{\mathsf{p}}
\newcommand{\pLaplaceConj}{\overline{\pLaplace}}
\newcommand{\Vbold}{\mathbf{V}}

\newcommand{\appropto}{\mathrel{\vcenter{
  \offinterlineskip\halign{\hfil$##$\cr
    \propto\cr\noalign{\kern2pt}\sim\cr\noalign{\kern-2pt}}}}}

\newcommand{\Pbold}{\mathbf{P}}
\newcommand{\fpeak}{f_{peak}}
\newcommand{\Bcenter}{\B_{peak}}
\newcommand{\phiaccum}{\phi_{accum}}
\newcommand{\BWndBbeta}{\text{BW}_{n,\B}}
\newcommand{\BWndBf}{\text{BW}_{n,f}}
\newcommand{\CFx}{\text{CF}(x)}

\usepackage{diagbox}

\newcommand{\removeInShortVer}[1]{}

\newcommand{\figVer}{PonlymanyParamsSharpVsPhysNewesterest.png}

\begin{document}

\title{Characteristics-Based Design of Generalized-Exponent Bandpass Filters}

\author{Samiya~A~Alkhairy

\thanks{S. Alkhairy is with MIT, Cambridge, MA, 02139, USA. e-mail: samiya@mit.edu, samiya@alum.mit.edu}}



\markboth{Alkhairy - Characteristics-Based Design of Generalized-Exponent Bandpass Filters}{Alkhairy - Characteristics-Based Design of Generalized-Exponent Bandpass Filters}
\maketitle

\begin{abstract}
We develop characteristics-based filter design methods for a class of IIR bandpass filters, which we refer to as Generalized-Exponent Filters (GEFs) and that are represented as second-order filters raised to non-unitary exponents. GEFs have a peak, are effectively linear phase, and are useful for seismic signal phase-picking, cochlear implants, and equalizers. 
The native frequency-domain specifications for GEFs are not on given frequency responses but rather on filter characteristics such as peak frequency, bandwidth, and group delay. Our characteristics-based method for filter design accommodates direct specification of a trio of frequency-domain characteristics from amongst the peak frequency, convexity, ndB quality factors, equivalent rectangular bandwidth, maximum group delay, and phase accumulation. 
We achieve this by deriving filter parameterizations in terms of sets of filter characteristics which involves deriving closed-form analytic expressions mapping sets of filter characteristics to the original filter constants by making sharp-filter approximations.
This results in parameterizations for GEFs including ones with simultaneous specification of magnitude-based characteristics and phase-based characteristics (e.g. bandwidths and group delays). This in turn enables designing sharply tuned filters without significant group delay, and simultaneous control over frequency selectivity and synchronization which is important in designing filterbanks. Our filter design methods with direct control over characteristics may also be utilized beyond static filter design for higher-order variable bandpass filter design and may be useful for characteristics-based adaptive filtering. Our methods are inherently stable, highly accurate in meeting strict specifications on desired characteristics, simple, and computationally efficient. The methods extend to the design of related bandpass and multiband filters.
\end{abstract}



\IEEEpeerreviewmaketitle


\section{Introduction}
\label{s:intro}
Our goal is to develop IIR filter design methods that enable directly designing filters from desired sets of frequency-domain filter characteristics (e.g. cutoff frequency, rolloff, peak frequency, group delay, and quality factors) which are often-times the native specifications for various classes of filters. This is in contrast to filter design methods that require specifications on the frequency response or its magnitude or amplitude and tolerances. 

We develop the characteristics-based filter design methods for a class of IIR filters in their bandpass region of operation. The methods for filters design are not only based on native specifications on characteristics, but also allows for simultaneously specifying a trio of frequency-domain characteristics beyond the peak magnitude, including mixed magnitude-based characteristics and phase-based characteristics (e.g. quality factor and group delay), and enables the design of sharp filters with minimal delay. The methods are simple, highly accurate, computationally efficient, and inherently stable, and are also useful for the design of related static bandpass and multiband filters and filterbanks, and variable digital bandpass filters. The methods may be used for both analog and digital filter design, but our interest is primarily is in digital filters due to the multitude of applications.

In the remainder of this introduction we (a) introduce existing methods for filter design, and (b) present the motivation and objective for developing our characteristics-based  design methods. In section \ref{s:filterClass}, we present the general form for the main class of filters for which we develop our characteristics-based filter design methods along with applications for that class of filters. In section \ref{s:considerationsmethodology}, we discuss considerations that we take into account when developing and testing our filter design methods, and introduce relevant filter characteristics. In section \ref{s:modelExpressions}, we introduce normalized frequencies, and the transfer functions of the various classes of filters for which we develop our characteristics-based filter design methods.  In section \ref{s:tunabilityApproach}, we describe our detailed approach towards developing methods for characteristics-based filter design. In sections \ref{s:tunabilityOverview}-\ref{s:multiband}, we develop, verify, and test our filter design methods. Finally, we discuss properties of our methods in section \ref{s:features}, and provide conclusions and future directions in section \ref{s:conclusion}.
 
\subsection{Existing Methods for the Design of Static Digital IIR Filters}
\label{s:existingMethods}

Filter design methods are generally developed in the frequency domain, but may also be in the time domain \cite{galvez2015time}. With the exception of characteristics-based methods for design of first and second order filters, existing methods are based on specifications on frequency-responses. These frequency-response based design paradigms for digital filters aim to estimate filter coefficients that meet certain criteria - e.g. maximally flat response in the passband, equiripples, specified cut-off frequencies, frequency response; and the filter design problem is formulated to be optimal with respect to certain criteria. 

There are several methods for digital filter design, which may generally\footnote{This generalization is mostly true, though  there exist smaller subsets of other approaches. For instance, we note that while FIR filters are generally designed using optimization techniques there exist classical approaches based on windowing methods - see \cite[Chapter~7]{oppenheim1999discrete}\removeInShortVer{,  \cite[Chapter~8,9]{antoniou2006digital},\cite{rader1967digital}}. Another example is the more recent form of polynomial-based filter design of half-band filters (with certain symmetries) which are used in decimation and interpolation \cite{cho2021design}.}
be categorized as (A) iterative optimization-based design (for both FIR and IIR filters) \cite{milic2022robust}, and (B) direct design based on analog prototype filters (for IIR filters). Recent filter design methods also include those that utilize machine learning such as \cite{pepe2022deep}, which are related to the first category. \removeInShortVer{Digital filter design of IIR filters is classically done based on analog prototype filters - such as Butterworth and Chebyshev, that are stable and have known optimal behavior (e.g. optimal conditions on ripples in the pass/stopband and flatness) and then converted to a digital filter using techniques such as bilinear transforms. In addition to the analog-prototype-based method for designing IIR filters, there also exist optimization-based methods for solving for IIR filter coefficients. Several iterative methods originally developed for FIR filters have been extended to formulate and solve optimization-based filter design methods for IIR filters - though the problem is more involved due to issues of stability and pole-zero placement. FIR filter design is typically carried out by solving an optimization problem to find coefficients to minimize some cost function on the difference between the target and actual frequency response specification.} For category (A), the native specifications may be on the frequency response itself, or alternatively be on characteristics such as cut-off frequency, roll-off, or bandwidths, which then must be converted to a set of specifications on frequency response and tolerance. In order to use such methods, a filter form and order are specified, and a desired specification of a set length is supplied. This specification may be on magnitude, amplitude \cite{nordebo2000use}, or the complex frequency response \cite{alkhairy1993design}. The objective function is defined based on certain criteria - e.g. minimax and least squares \cite{lu2017design}, then solved for filter coefficients.


\subsection{Motivation and Objective}
\label{s:motivationGoal}

In many cases, the native specifications for filter design are in terms of filter characteristics (e.g. cutoff frequency, roll-off, peak frequency, bandwidth, and/or group delay for linear-phase filters) rather than in terms of the frequency response (or its magnitude or amplitude) that is specified for existing filter design methods. This motivates the development of a third category of design methods for various classes of filters that are directly based on native specifications of filter characteristics. Towards this end, we develop characteristics-based filter design methods for a class of bandpass IIR filters which we refer to as Generalized Exponent Filters (GEFs) that display a clear peak, are effectively linear-phase, and are useful for signal processing applications that would especially benefit from characteristics-based design methods.

\section{Class of Filters}
\label{s:filterClass}
Here we describe the main class of filters for which we derive our characteristics-based filter design methods. We also present applications that use such filters, and provide our rationale for focusing on this class of filters in this paper that introduces characteristics-based filter design. 
\subsection{General Representation}

The class of bandpass filters for which we develop characteristics-based filter design methods has a transfer function of the form, 

\small
\begin{equation}
    H(\sLaplace) = C \big((\sLaplace-\pLaplace)(\sLaplace-\pLaplaceConj) \big)^{-\Bu} \;,
    \label{eq:filterForm}
\end{equation}
\normalsize

in $\sLaplace$ where $-\Re\{\pLaplace\}, \Im\{\pLaplace\} \in \mathbb{R^+}$ and $\Bu \in \mathbb{Z^+}$ resulting in an all-poles stable filter with a single repeated pair of complex conjugate poles in the left half-plane \footnote{The constant $\Bu$, is in fact not limited to positive integer values and may take on positive rational values. However, the stability, causality, and reliability of the non-integer exponent case is not discussed here and is instead addressed separately \cite{paperB2}.}. Hence, stability, which is difficult to achieve for filters in general, is easily satisfied by controlling only a single pole pair. The additional degree of freedom compared to second order filters (i.e. the exponent applied to the second order base filter) allows for having large quality factors without the need for correspondingly large delays as will be shown later in this paper.

We refer to these filters in the bandpass region of operation as Generalized Exponent Filters (GEFs). GEFs exhibit a distinct peak (rather than flat passbands) in the magnitude of the frequency response and are functionally linear-phase as the group delay in the bandpass region is constant as later demonstrated in this paper. We are unaware of any existing filter design methods that have been developed to specially take into consideration the properties of this class of filters (or of related classes of multi-exponent filters) and especially the fact that the native specification is on filter characteristics rather than frequency responses. Later in this paper, we demonstrate that the filter design methods we develop for GEFs are also suitable for some related classes of bandpass and multiband filters.

\subsection{Filter Applications}

Filters that can be represented by (\removeInShortVer{equation }\ref{eq:filterForm}) occur in a wide variety of systems - the simplest of which are second order systems such as certain configurations of RLC circuits which have been studied extensively. Filterbanks composed of such filters and similar ones (e.g. gammatone filters) are useful for a wide range of applications including phase-picking from micro-seismic recordings \cite{jiang2020automatic}, \removeInShortVer{spike-based pattern recognition and seismic event detection \cite{dibazar2010statistical},} fault diagnostics - including those used in nondestructive testing  \cite{abdul2020hybrid}, rainbow sensors \cite{karlos2020cochlea}, underwater sound classification and target feature extraction \cite{zhang2018underwater}, cochlear implants \cite{harczos2012making}\removeInShortVer{, \cite{cosentino2013cochlear},  \cite{yang2014rm}}, hearing aids \cite{sokolova2022real}, speech feature extraction and speech recognition \cite{alias2016review}. Associated multiband filters may also be used for a wide range of applications such as auditory equalizers \cite{jankovic2015design} and denoising signals from mechanical systems with known properties \cite{feng2021filter}.

For certain ranges of values of the filter constants and their variation within a filterbank, the GEFs mimic auditory/cochlear signal processing - specifically in response to stimuli at low sound levels \cite{alkhairy2019analytic}. We will refer to GEFs with these ranges of filter constant values as Auditory Filters/Filterbanks (AFs). For several of the diverse applications, it is desirable to design the filterbanks such that they mimic auditory signal processing and behavior - e.g. \cite{deepak2021convolutional}. Indeed, certain multiplexing filter topologies are even inspired by the cochlea \cite{galbraith2008cochlea}. Consequently, we especially test the accuracy of our methods for various ranges of values that apply for AFs.


\subsection{Why GEFs}

We chose GEFs as the class of bandpass filters for which we introduce the characteristics-based design methods. This choice is primarily made for two reasons: (a) For this class of filters, the native specifications are on characteristics rather than on frequency responses, and (b) GEFs - as will become apparent later in this paper, are highly controllable in the sense that we can design filters with large quality factors but small group delays while still retaining a small number of degrees of freedom unlike typical higher-order filters. The small number of filter constants and the use of a second order filter as a base filter results in the ability to build intuition,  more easily understand the filter behavior, and results in tractable derivations. Other reasons for choosing GEFs include its utility for a diverse array of applications, its simple stability analysis and control, and its effectively linear-phase behavior which limits signal distortion.

\section{Considerations for Filter Design and Choice of Domain}
\label{s:considerationsmethodology}
Our goal is to design filters directly based on native specifications of filter characteristics rather than specifications on frequency responses. We mainly do so for the class of filters introduced in section \ref{s:filterClass}. In this section, we (a) discuss considerations for filter design which we must keep in mind when developing and assessing our methods, then (b) describe the domain in which we develop our filter design methods, and finally (c) provide an overview of the remaining sections in this paper so that readers of varying interests may focus on sections that are of greatest relevance to them.

\subsection{Considerations for the Development of Methods}
\label{s:considerations}
\removeInShortVer{ Our filter design method diverges from the two primary categories of methods for the design of IIR filters, (A) iterative optimization-based methods, and (B) analog-prototype based methods. Consequently, }

Here we discuss fundamental considerations for filter design that provide the context in which we develop and assess the methods introduced in this paper. Specifically, we first discuss considerations for the design of static filters, then proceed to discuss additional considerations for the design of filters with changeable coefficients, and finally, additional for the design of filterbanks. After developing the filter design methods and performing verification and testing in later sections, we return to these considerations and evaluate our filter design methods in section \ref{s:features}.

\subsubsection{Considerations for Static Filter Design}
In developing methods for digital filter design, one must consider the following: the family and function of filters; the form of desired specifications for the filter design method; constraints; and desired features. The desired features may include causal-stability guarantees or checks, simplicity, computational efficiency, optimal or local minima, control over both magnitude and phase (or their characteristics), control over multiple characteristics simultaneously (e.g. cutoff and roll-off), and accuracy in achieving desired specifications.

The most fundamental aspects are: considering what the native set of specifications are in the first place (e.g. specifications on characteristics such as bandwidths or specifications on frequency responses over a length of frequencies), and the ability to directly use these native specifications for filter design - rather than needing additional stages of generating an alternate set of specifications which are then in turn used to design the filter. The set of native specifications is determined based on the class of filters and its applications. We may evaluate methods for static filter design based on the aforementioned considerations and features.

\subsubsection{Additional Considerations for the Design of Filters with Changeable Coefficients}
The considerations for filter design methods increase as one moves from single static filters to other problems such as the case of filters with changeable coefficients. This occurs in the design of adaptive filters and variable digital filters for which retaining these criteria across evolutions becomes of paramount importance \cite{koshita2017variable, yu2011mixed}. Additionally, in some instances where the coefficients vary with time, it is desirable that certain frequency-domain characteristics of the filters (e.g. peak frequency and delay) are controlled so that they can be varied in a smooth manner and that distortions are minimized. 

In the case of variable digital filters \footnote{alternatively referred to as changeable-parameter / variable / tunable / adjustable / programmable filters \cite{stoyanov1997variable}.} and their applications - including characteristics-based adaptive filtering \cite{nehorai1985minimal, koshita2018recent, yamaguchi2004adaptive}, the objective is to tune a previously-designed filter based on one or more filter characteristics such as the cut-off frequency or peak frequency - thereby necessitating parameterizations in terms of such characteristics \footnote{In the variable digital filter literature, the frequency-domain characteristics such as peak frequency and bandwidths are often referred to as `spectral parameters'.}. The advantage of applying variable digital filters to characteristic-based adaptive filtering is especially noticeable in the case of a single tunable characteristic - e.g. cutoff frequency, where the cost function notably becomes unimodal and a global optimum is achieved.

\subsubsection{Additional Considerations for the Design of Filterbanks}

Another formulation in which additional filter design features are required are filterbanks. In this case, we want to specify the peak frequencies and bandwidths of constitutive filters and limit cross-talk, and also want to specify the delay across filters for requirements on minimum delay or synchronization. This, too, would be most easily achieved if one is able to design filters directly based on the characteristics.

\subsubsection{Criteria and Assessment for Proposed Methods}
In developing our methods for filter design, we aim to fulfill the majority of the aforementioned criteria (e.g. direct control over native specifications, high accuracy in meeting specifications, simplicity, computational efficiency, simultaneous control over characteristics). We return to these considerations and assess our methods in their context later in this paper (section \ref{s:features}). We note that - with the exceptions of methods for first and second order filters, each of the existing filter design methods only fulfills a subset of these considerations.

\subsection{Domain and Types  of Filter Characteristics}
\label{s:proposed}

\subsubsection{Domain for Method Development}

To develop our filter design method, we derive characteristics-based parameterization in continuous frequency regardless of whether the filter is used in an analog or digital capacity. We do so as continuous frequency is the most suitable domain for deriving expressions for \textit{frequency}-based characteristics. For use in the digital domain, the filter must then be converted to digital filters using analog-to-digital techniques (such as bilinear transforms \cite{pei2008fractional} and numerical-integration-based methods \cite{paperB2}) bearing in mind which desired features should be retained (e.g. stability, accuracy, and limited frequency warping). In this regard, our characteristics-based filter design method is similar to analog-prototype-based methods for the design of digital filters.


\subsubsection{Sets of Characteristics Used for Filter Design}

As a result of choosing the continuous frequency domain, the characteristics we use for filter design are naturally defined in that domain. The proposed filter design method results in the ability to design the GEFs - as well as related bandpass filters, multiband filters, and filterbanks, by specifying values for viable sets of three frequency-domain filter characteristics, $\Psi$. These trio of filter characteristics may include: peak frequency, 3 dB bandwidth, 10 dB bandwidth, equivalent rectangular bandwidth, phase accumulation, group delay at the peak frequency, as well as convexity - a characteristic we introduce in this paper which is a peak-centric measure of sharpness of tuning.


\subsection{Content}

Here we provide additional details regarding the contents and organization of this paper, and introduce relevant notation. We also provide a guide so that readers from various backgrounds may choose to focus on sections that are of greatest relevance to them. 

As we build on concepts of characteristics-based design and parameterizations from commonly-known second-order filters, we review these filters in appendix \ref{s:secondOrderToy}. In section \ref{s:modelExpressions}, we  provide details regarding GEFs and their transfer functions $\Pbold(s; \Theta)$ that are parameterized by filter constants $\Theta$. We also introduce related classes of filters which can also directly be designed using the characteristics-based filter design methods we develop in this paper.

In sections \ref{s:tunabilityApproach} and  \ref{s:tunabilityOverview}, we develop the filter design paradigm by mapping the problem into one of expressing filter constants, $\Theta$ - the pole and filter exponent, in terms of desired filter characteristics, $\Psi$, such as sets that may include peak frequency, quality factor, phase accumulation, and maximum group delay. This results in expressions for filter constants in terms of characteristics, $\Theta = g(\Psi)$, as in Table. \ref{tab:char2param}. We may use these expressions to parameterize GEFs using a subset of these filter characteristics, $\Pbold(s; \Psi)$, as will be exemplified in (\ref{eq:parametizationNandPhiaccum}) and (\ref{eq:parametizationNandS}) of section \ref{s:tunabilityOverview}. We verify our derivation using the subclass of second order filters.

We demonstrate the high accuracy of our characteristics-based filter design approach for GEFs as well as related bandpass and multiband filters in sections \ref{s:evaluation} and \ref{s:multiband}. We do so using various ranges of desired filter characteristics. We describe features of the methods in section \ref{s:features}, referring back to  the considerations we set in section \ref{s:considerations}. We provide conclusions and discuss future directions in section \ref{s:conclusion}.

Readers who are only interested in applying the characteristics-based methods for GEFS or related filters, may read section \ref{s:modelExpressions}, the skip sections \ref{s:tunabilityApproach}-\ref{s:multiband} with the exception of Table. \ref{tab:char2param} and section \ref{s:exampleparams}, the proceed to sections \ref{s:features}-\ref{s:conclusion}.
Readers who are interested in understanding and assessing the filter design methods developed here, and those who are interested in developing, verifying and testing characteristics-based design methods for other classes of filters are encouraged to read all sections.


\section{Expressions for GEFs and Related Filters}
\label{s:modelExpressions}


In this section, we present the relevant filter transfer functions that we later parameterize with sets of filter characteristics used for filter design. In order to develop our characteristic-based filter / filterbank design method, we first introduce the GEF expressions in more detail and simplify the parameterization by using normalized frequencies \footnote{We use notation from AFs introduced in \cite{alkhairy2019analytic} which relates the filters to expressions based on a mechanistic model of the cochlea.}. We also introduce related bandpass filters for which the GEF-based filter design methods apply quite directly and accurately.



\subsection{GEF Transfer Functions}

Instead of considering filters in angular frequency, $\w$, we consider them in terms of normalized frequency, 

\small
\begin{equation}
    \B \triangleq \frac{\w}{\w_{peak}} \;, 
\end{equation}
\normalsize

similar to \removeInShortVer{equation }(\ref{eq:soswwn}) \footnote{We use normalized frequency, $\B$, for two reasons: (a) In the case of filterbanks constituted of constant-Q filters (which all have the same quality factors but different peak frequencies and bandwidths), the filter representation in terms of $\B$ requires the same set of filter constant values across all filters; and (b) AFs (GEFs that mimic auditory signal processing in the cochlea) can be associated with certain cochlear models in which the peak frequency and bandwidths vary along the cochlear length (vary with filters in the auditory filterbank) in such as way, that locally, the filters can be considered a function of $\B$ (or locally, constant-Q).}. As we express the filters in terms of normalized  frequency, $\B$, rather than regular frequency, we define the filter characteristics in $\B$ (in section \ref{s:tunabilityOverview}).  

Due to using $\B$, we deal with a normalized version of $\sLaplace$. This results in an $s$ that - when purely imaginary, may be expressed as,

\small
\begin{equation}
    s(\B) = i\B \;.
\end{equation}
\normalsize

As specifying the gain constant is trivial, we consider the GEF transfer function to be $\Pbold(\B)$,

\small
\begin{equation}
\Pbold(\B; \Theta) = \Big((s-p)(s-\bar{p})\Big)^{-\Bu} \;,
    \label{eq:Pbold}
\end{equation}
\normalsize

where $p = i\bp - \Ap$ and  $\pconj = -i\bp - \Ap$. The filter constants in $\Theta = [\Ap, \Bu, \bp]$ take on positive real values. 

Clearly, for the case of an integer-exponent $\Bu$, \removeInShortVer{equation }(\ref{eq:Pbold}) represents rational transfer functions of IIR filters that are stable and causal. These filters may be considered a cascade of second order filters where all constitutive filters have the same peak frequency, thereby resulting in a bandpass filter with a single peak. The multi-exponent form allows for higher order filters compared to second order filters - and therefore the potential to design filters large quality factors but minimal delay,  while limiting the number of additional filter constants to one. 

\removeInShortVer{
In the next section, our derivation of expressions for filter characteristics - towards developing our filter design method, requires that the magnitude (in dB) and phase of $\Pbold$ are expressed separately - which may be easily obtained from the real and imaginary parts of $\log(\Pbold)$. Consequently, we present a variable, $\kB(\B)$, from which $\Pbold$ was derived in \cite{alkhairy2019analytic} \footnote{In the cochlear model from which we obtain the AF expressions, $k(\B) = \kB(\B) \frac{\B}{l}$ corresponds to the differential pressure wavenumber and is considered to contain important mechanistic information regarding how we process sounds.}: 

\small
\begin{equation}
    \kB(\B) = i \frac{d\log(P)}{d\B} \;.
\end{equation}
\normalsize

The expression for $\kB(\B)$ is, 

\small
\begin{equation}
    \kB(\B) = 2\Bu \frac{s+\Ap}{(s-p)(s-\pconj)} = \Bu \bigg( \frac{1}{s-p} + \frac{1}{s-\pconj} \bigg) \;,
    \label{eq:wavenum}
\end{equation}
\normalsize

where we have included the partial fraction decomposition which greatly simplifies our derivation in the following sections.
}


\subsection{GEF Filterbanks}

We may use (\ref{eq:Pbold}) to develop the design methods for single filters as described in the following section. However, as many applications use GEFs in filterbank configurations, we also discuss the formulation of filterbanks here. Readers who are only interested in filters and not filterbanks may skip forward to the next section. 

In the particular case of filterbanks (in parallel configurations), we must set the peak frequency of each constitutive filter based on some predetermined designed variation within the filterbank. For instance, for filterbanks that mimic auditory signal processing - as is desired in a wide range of applications, it is desirable to have the peak frequency for the constitutive filters follow the variation of peak frequencies along the length of a cochlea. In this case, we may set $\bp = 1$ and express $\B$ as,

\small
\begin{equation}
    \B = \frac{f}{\CFx} \;,
    \label{eq:BetaWithCFx}
\end{equation}
\normalsize

where $\CFx$ is a characteristic frequency map which takes the form, $\CFx = \text{CF}(0) \expn{-\frac{x}{l}}$. $x$ is location along the length of a cochlea, and $l$ is a space constant. Consequently, a filter in a parallel filterbank configuration has a unique peak frequency or, equivalently, a corresponding unique location along the length of a fictitious cochlea - thereby allowing us to refer to a filter by its CF or by its $x$. 

For GEFs in a filterbank configuration, the gain constant becomes relevant and the filters are represented as,
\small
\begin{equation}
    \frac{P(x,\w)}{v_{st}(\w)}  = C(x) \Pbold(\B(x,\w); \Theta(x)) = C(x) \Big((s-p(x))(s-\bar{p}(x))\Big)^{-\Bu(x)} \;,
    \label{eq:Ptovst}
\end{equation}
\normalsize

with the outputs of the filterbank being $P(x,\w)$ for a given input, $v_{st}(\w)$ \footnote{In the cochlear model corresponding to AFs, the filter/response variable $P$ corresponds to differential pressure right across the organ of Corti (OoC), $P(x,\w)$, and the input, $v_{st}(\w)$, is the stapes velocity.}. For constant-Q filters, the filter constants are independent of $x$,

\small
\begin{equation}
    \frac{P(x,\w)}{v_{st}(\w)}  = C(x) \Pbold(\B(x,\w); \Theta) = C(x) \Big((s-p)(s-\bar{p})\Big)^{-\Bu} \;.
    \label{eq:XX}
\end{equation}
\normalsize

\subsection{Related Filters}

As most of the filter characteristics we consider in this paper are peak-centric, our filter design methods also approximately apply to a class of filters that are related to GEFs and which have a transfer function,

\small
\begin{equation}
    \Vbold_{(\Ap; \Bu)}(s) \propto (s+\Ap)\Big((s-p)(s-\bar{p})\Big)^{-\Bu} \;,
    \label{eq:V}
\end{equation}
\normalsize

which has a single zero at $-\Ap$ and the pole-pair is repeated $\Bu$ times \footnote{This is related to the velocity of the Organ of Corti partition in the cochlear model, where it is of the form $\Vbold_{(\Ap; \Bu+1)}$ and is tied to other model variables. It also bears some similarities to the One Zero Gammatone Filter.}. We henceforth refer to the above class of filters as $\Vbold$.

We note that $\Vbold$ behaves most similarly to $\Pbold$ when the filter is sharply tuned, and we later demonstrate (in section \ref{s:evaluation}) that the design methods developed here also apply directly to $\Vbold$ and are very accurate for reasonable values of $\Ap$.

The peak-centric characteristics-based design methods developed based on GEFs also approximately apply to other classes of filters such as, 
\small
\begin{equation}
    \mathcal{Q}_{(0;c;\Bu)}(s) \propto s^c \big((s-p)(s-\pconj) \big)^{-\Bu} , \quad \textrm{with } c < \Bu  \;.
    \label{eq:filterFormZero}
\end{equation}
\normalsize

We can design such filters based on the characteristics-based parameterization derived in this paper for GEFs provided that the zero is far enough from the peak frequency. Additionally, the characteristics-based design paradigm presented here for single-band filters can be used for multi-band filters with distinct peaks as we demonstrate later in this paper (section \ref{s:multiband}). Consequently, the filter design approach developed here can be used for a wider variety of filters than GEFs (\removeInShortVer{equation }\ref{eq:filterForm}) - which itself may be used for a wide range of applications.


\section{Approach for Developing Characteristics-Based Design Methods}
\label{s:tunabilityApproach}

In developing our methods for characteristics-based parameterization and design for GEFs, it is not possible to simply extrapolate and use the equations for characteristics of the related second order systems (or equivalently, place poles based on the quality factor of the associated second-order system which is distinct from the quality factor of a higher-order filter \footnote{In many representations, the quality factor or bandwidth of the second order base filter is used to parameterize the transfer function of higher-order filters resulting in inaccurate filter design as many users erroneously take it to indicate the quality factor of the filter itself - likely due to overusing notation \cite{lyon1996all}}). Instead, the characteristic-based parameterization must be properly derived from the expressions for the higher order filter itself. 

The filter expressions in the previous section are parameterized by the pole and filter order, or alternatively, by the filter constants $\Theta = [\Ap, \bp, \Bu]$. In order to develop our characteristics-based filter design method, we aim to instead parameterize the GEFs in terms of a set of desired frequency-domain filter characteristics, $\Psi$, which describes the behavior of the  magnitude and/or phase of the transfer functions. These filter characteristics include bandwidths, quality factors, maximum group delays, peak frequencies, and phase accumulations. \removeInShortVer{In this paper, we are not interested in filter design based on time-domain characteristics - whether those based on step responses or impulse responses \cite{charaziak2023estimating}, which require a separate derivation.}

\begin{figure*}[htbp]
    \centering
    \includegraphics[width=1\linewidth]{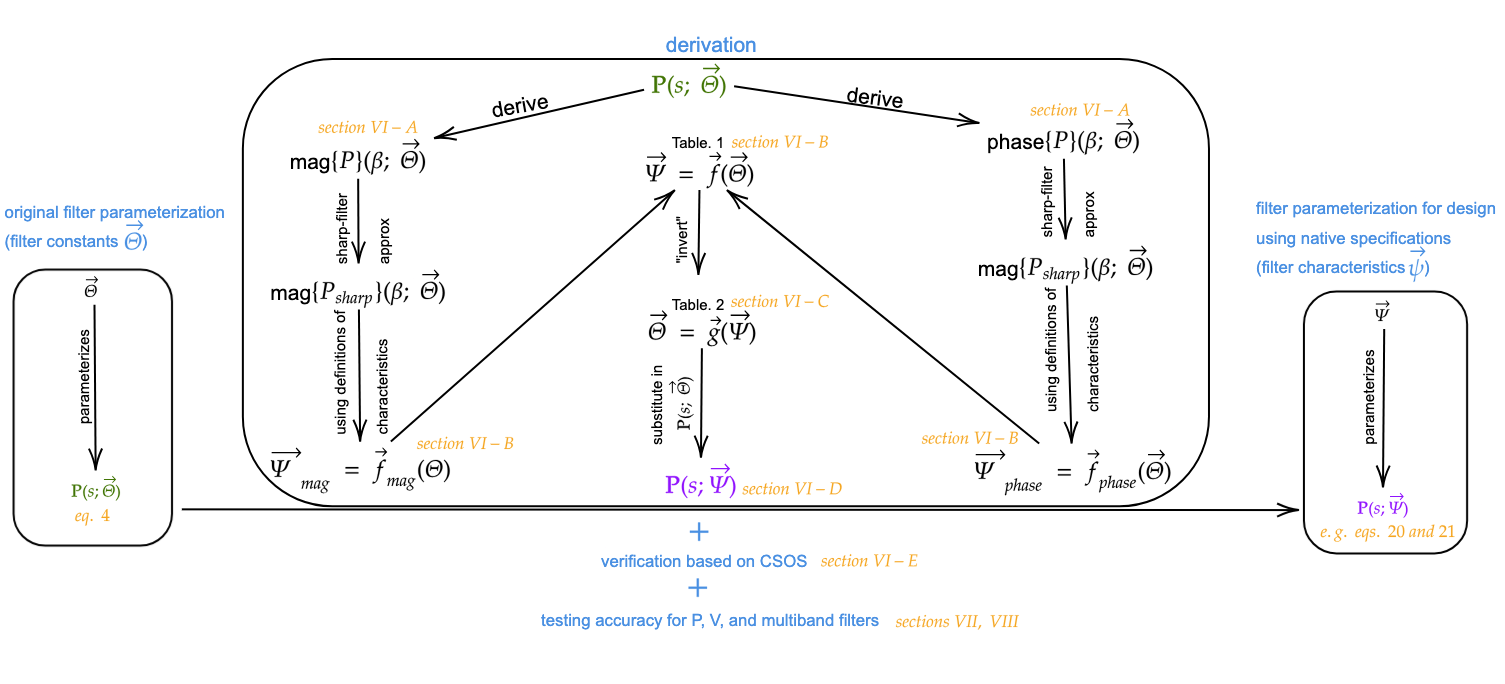}
    \caption[]{\textbf{Schematic diagram describing the approach for developing the characteristics-based filter design methods and characteristics-based filter parameterizations:} We describe this approach in detail in the main text of section \ref{s:tunabilityApproach}. We refer to relevant sections, tables, and figures in the schematic so that the reader may easily refer to these. We start with the transfer function in green which is parameterized by filter constants and end with the transfer function in purple which is parameterized by filter characteristics that are the native specifications used for filter design. We have used the arrow notation to emphasize the vector nature of the filter constants and filter characteristics - though we avoid doing so in the main text for simplicity.}
    \label{fig:approachschematic}
\end{figure*}

To develop our method for characteristics-based design of GEFs and related filters, we perform the following steps  as schematized in Fig. \ref{fig:approachschematic}:

\begin{enumerate}
    \item First derive \textit{separate} expressions for the magnitude and phase of GEFs parameterized by $\Theta$ - i.e. $\level{\Pbold(\B; \Theta)}$ and $\phase{\Pbold(\B; \Theta)}$. This is because each of the filter characteristics, $\Psi$, is defined based on \textit{either} the magnitude of the transfer function (e.g. peak frequencies, quality factors) \textit{or} its phase (e.g. maximum group delays, phase accumulations) alone.
    \item Simplify the aforementioned expressions to arrive at $\level{\Pbold_{sharp}(\B; \Theta)}$ and $\phase{\Pbold_{sharp}(\B; \Theta)}$ using a sharp-filter approximation. This is done so that the following steps in the derivation are more tractable.
    \item Derive expressions for the magnitude-based and phase-based filter characteristics $\Psi$ in terms of the filter constants $\Theta$ - i.e. derive expressions $\Psi = f(\Theta)$ using the simplified expressions from the previous step.
    \item Invert the expressions, $\Psi = f(\Theta)$, to derive expressions for the filter constants in terms of sets of filter characteristics - i.e. $\Theta = g(\Psi)$ as will be seen in Table. \ref{tab:char2param}. Any such sets may be used for characteristics-based design.
    \item Finally, we may express GEFs as parameterized by any chosen set of desired filter characteristics (rather than filter constants), $\Pbold(\B; \Psi)$ which is specified for filter design. This characteristics-based parameterization of the GEFs transfer function will be exemplified in (\ref{eq:parametizationNandPhiaccum}) and (\ref{eq:parametizationNandS}).
    \item Towards digital implementations, the GEFs that are parameterized by $\Psi$ are converted to their discrete frequency counterparts using one of the existing techniques (e.g. \cite{al2007novel, nelatury2007additional, goswami2021extended, paarmann2006mapping, paperB2}) bearing in mind desired properties including retaining stability and limiting the reduction in accuracy introduced due the conversion \cite{van2003digital}. This is similar to the final stage in analytic analog-prototype methods for digital IIR filter design which is extensively studied in classical literature and hence not discussed here. Various  architectures for similar classes of filters have already been designed \cite{van2003digital, katsiamis2007practical}.
\end{enumerate}


As previous mentioned, the filter design paradigm developed in this paper may be used not only for GEFs, but also other bandpass filters that can be approximated as GEFs near the peak - such as $\Vbold$ and filters described by \removeInShortVer{equation }(\ref{eq:filterFormZero}), or dual-band and multi-band filters that can be constructed based on the single-band filters designed here. Consequently, it is desirable to test the accuracy of the filter design method for each of these classes individually. We perform these tests later in this paper for $\Pbold$ and $\Vbold$ and to some extent for multi-band filters constructed from GEFs.

\section{Developing Methods for Characteristics-Based Design and Parameterization}
\label{s:tunabilityOverview}

In this section, we develop the characteristics-based filter design methods which  enable us to design GEFs with desired frequency-domain filter characteristics that are the native specifications for such filters. We develop these methods by following the approach we introduced in the previous section and in Fig. \ref{fig:approachschematic} which we encourage the reader to periodically refer back to as they read through this section. Those who are only interested in the final expressions and parameterizations used for filter design - rather than the detailed derivation in this section, need only focus on (a) the parts of subsection \ref{s:tunabilityCharsFxnOfParams} which introduce the filter characteristics, followed by (b) Tables. \ref{tab:charInTermsOfParams} and \ref{tab:char2param} as well as (\ref{eq:parametizationNandPhiaccum}) and (\ref{eq:parametizationNandS}) for the characteristics-based parameterizations used for filter design.

\subsection{Expressions for Magnitude and Phase}

As stated in the previous section and shown in Fig. \ref{fig:approachschematic}, the first two steps towards developing our filter design methods are to (a) derive separate expressions for the magnitude and phase of GEFs ($\Pbold$) as parameterized by the constants $\Theta$, and to (b) simplify these expressions using a sharp-filter approximation which ultimately makes the derivation of the characteristics-based parameterization much more tractable.

\removeInShortVer{This is most easily done by integrating the separate expressions for the real and imaginary parts of the $\kB$ \footnote{Integrating the real and imaginary parts of $\kB$ simplifies our derivation immensely due to the fact that we are able to express $\kB$ easily using partial fraction decomposition \removeInShortVer{equation }(\ref{eq:wavenum}) which leads to simpler individual terms}, due to the relationships,

\small
\begin{equation}
\frac{d}{d\B} \level{\Pbold} = \frac{20}{\log(10)} \Im\{\kB\} \;,
\end{equation}
\normalsize

and,

\small
\begin{equation}
\frac{d}{d\B} \phase{\Pbold} =  -\Re\{\kB\} \;.
\end{equation}
\normalsize
}

This results in the following expressions for the magnitude and phase of $\Pbold$, where we use $\phase{.}$ to indicate phase in radians and $\level{.} = 20 \log_{10}(|.|)$ to indicate magnitude in dB - to distinguish it from $|\Pbold|$:

\small
\begin{equation}
\begin{split}
        \frac{\log(10)}{20} \level{\Pbold(\B)} & = -\frac{\Bu}{2} \log(\Ap^2 + (\B-\bp)^2) \\ 
        &   -\frac{\Bu}{2} \log(\Ap^2 + (\B+\bp)^2) \\ 
        \xrightarrow{\text{sharp-filter approx.}} & \frac{\log(10)}{20} \level{\Pbold_{sharp}(\B)} \\ & = -\frac{\Bu}{2} \log(\Ap^2 + (\B-\bp)^2) \;,
    \label{eq:Pmag}
    \end{split}
\end{equation}
\normalsize

and,

\small
\begin{equation}
\begin{split}
    \phase{\Pbold(\B)} & = -\Bu \tan^{-1}(\frac{\B-\bp}{\Ap}) -\Bu \tan^{-1}(\frac{\B+\bp}{\Ap})   \\ 
    \xrightarrow{\text{sharp-filter approx.}} & \phase{\Pbold_{sharp}(\B)} = -\Bu \tan^{-1}(\frac{\B-\bp}{\Ap}) \;.
    \label{eq:Pphase}
    \end{split}
\end{equation}
\normalsize

In the above equations, we have included sharp-filter approximations, $\level{\Pbold_{sharp}(\B)}$ and $\phase{\Pbold_{sharp}(\B)}$. For small $\Ap$, these expressions approximate $\level{\Pbold(\B)}$ and $\phase{\Pbold(\B)}$ quite well \footnote{with added shifts that do not depend on $\B$ and hence do not affect our derivations in the next sections}. Due to space considerations, we do have not included the rigorous derivation of the above approximations that we performed. However, we make the following notes here to build intuition for the reader. Let us first consider the phase: Near the peak, $\B \approx \bp$, and hence the argument in the inverse tangent function of the first term, $\B - \bp$, is close to zero whereas the argument of the second term, $\B + \bp$, is much larger.  Consequently, due to the shape of the inverse tangent function, the first terms varies appreciably with $\B$ (especially if $\Ap$ of the denominator is small) whereas the second term is approximately a constant with respect to $\B$. A similar argument may be made for the case of the magnitude with small $\Ap$ where the first term with $\B - \bp$ also varies appreciably with $\B$ whereas the second term is approximately just an offset independent of $\B$. 



\removeInShortVer{The condition for the sharp-filter approximation ($\Pbold(\B) \approx \Pbold_{sharp}(\B)$) is derived in appendix \ref{s:appendixSharp}.} Practically, the approximation is quite appropriate for filters with $\Ap < 0.2$ (when $\bp = 1$) which is the case for the majority of applications. While the sharp-filter approximation for GEFs is itself not physically realizable - as it has a transfer function $\Pbold_{sharp} = (s-p)^{-\Bu}$ (with an imaginary $p$), it enables us to derive expressions for the filter characteristics, $\Psi$, in terms of the filter constants, $\Theta$ and, consequently, parameterize the (physically realizable) GEFs ($\Pbold$) in terms of filter characteristics.

\begin{figure}[htbp!]
    \centering
    \includegraphics[width = \linewidth]{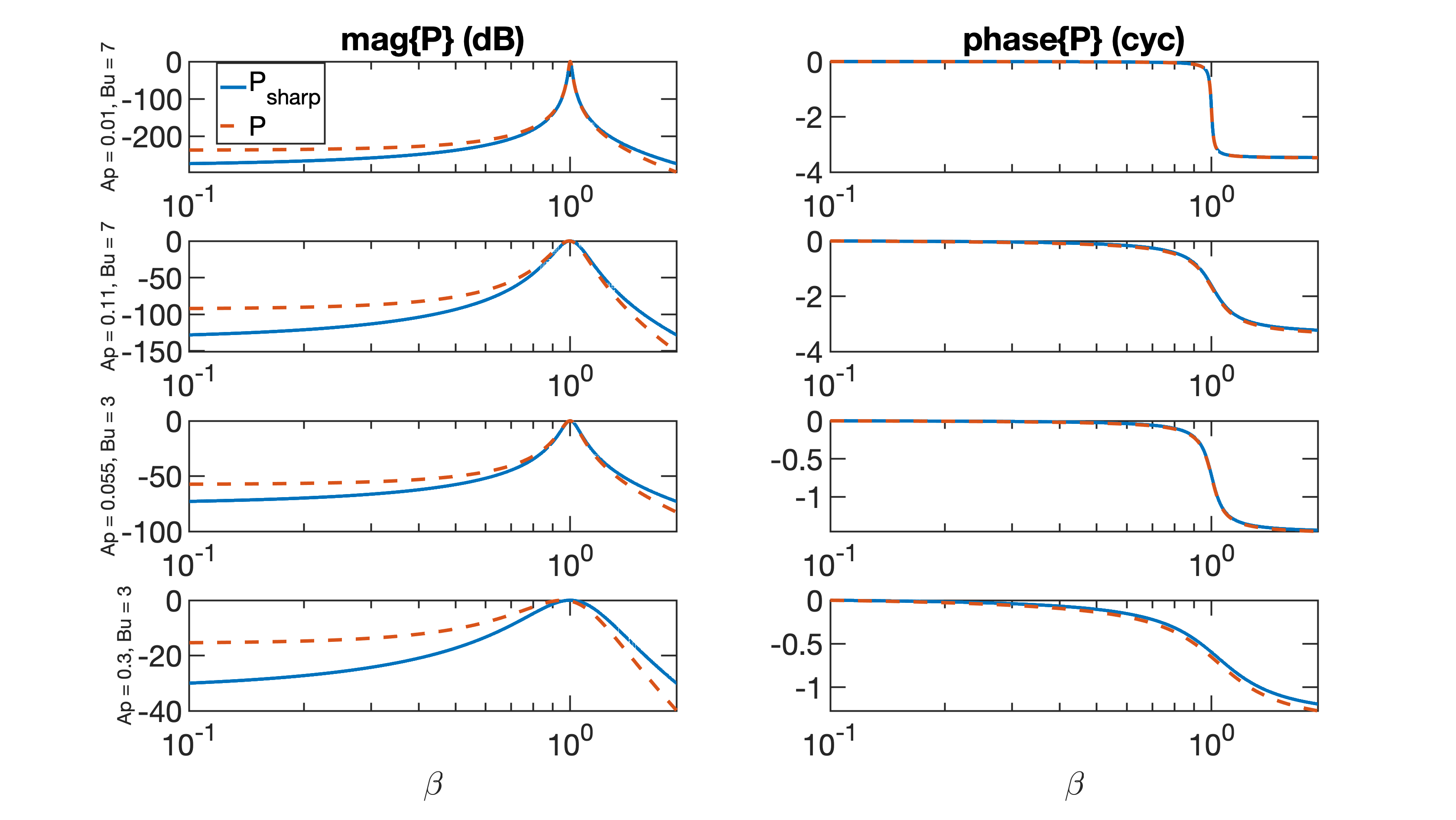} 
    \caption[]{\textbf{Validity of sharp-filter approximation and its dependence on filter constants:} The plots show the magnitude (left) and phase (right) of GEF/$\Pbold$ (dashed red lines) and its sharp-filter approximation, $\Pbold_{sharp}$ (solid blue lines) (normalized to peak magnitude and referenced to zero phase) \removeInShortVer{and real and imaginary parts of $\kB$} as a function of $\B$ generated using (\ref{eq:Pmag}) and (\ref{eq:Pphase}). Each row is generated using different values of the filter constants, $\Ap, \Bu$ with the last row \textit{not} fulfilling the sharp-filter condition $\Ap < 0.2$. For all plots, $\bp = 1$. The following may be observed: (a) the phase of $\Pbold$ and $\Pbold_{sharp}$ are almost equivalent, (b) the magnitudes are similar near the peak and especially so when the sharp-filter condition is fulfilled, and (c) the region of phase corresponding to an appreciable magnitude has a linear slope.}
    \label{fig:manySharpVsPhysP}
\end{figure}

In Fig. \ref{fig:manySharpVsPhysP}, we illustrate the parametric region of validity (in filter constant space) of the sharp-filter approximation for GEFs. We do so by plotting the magnitude and phase of GEF ($\Pbold$) and of its sharp-filter approximation, $\Pbold_{sharp}$\removeInShortVer{, as well as plotting the real and imaginary parts of $\kB$ and its sharp-filter approximation} for various values of filter constants. The sharp-filter approximation results in symmetric responses and does not preserve the asymmetry of $\Pbold$. For this reason, we do not include measures of asymmetry as filter characteristics we later use for filter design. In Fig. \ref{fig:manySharpVsPhysP}, we further demonstrate the following observations which may also be inferred from (\ref{eq:Pmag}) and (\ref{eq:Pphase}). For $\level{\Pbold}$\removeInShortVer{, $\Im\{\kB\}$}, the sharp-filter approximation is most valid near the peak of $\Pbold$ and for small $\Ap$. Consequently, magnitude-based filter characteristics derived from closest to the peak are most reliable. For instance, BW$_{15dB}$ is more reliable than BW$_{30dB}$ for filter design. As $\Ap$ increases or as we move away from the peak, the approximation deteriorates, as can be seen in the subplots. On the other hand, for  $\phase{\Pbold}$\removeInShortVer{, $\Re\{\kB\}$}, the sharp-filter approximation is a suitable approximation across $\B$. Consequently, the approximation is particularly good at preserving phase-based measures such as group delay. We note that for AFs (GEFs with filter constant values that mimic auditory signal processing), the sharp-filter approximation is a particularly good approximation in filters that mimic processing in the human base and apex \cite{alkhairy2017analytic, alkhairy2022cochlear}.

The sharp-filter approximations, $\level{\Pbold_{sharp}}$ and $\phase{\Pbold_{sharp}}$ make the filter more interpretable and are used to determine expressions for $\Psi = f(\Theta)$, and from it, $\Theta = g(\Psi)$, in order to then parameterize the transfer functions by filter characteristics which serves as the basis for our filter design paradigm.


\subsection{Frequency-Domain Filter Characteristics}
\label{s:tunabilityCharsFxnOfParams}

In the previous section, we derived separate expressions for $\level{\Pbold_{sharp}(\B; \Theta)}$ and  $\phase{\Pbold_{sharp}(\B; \Theta)}$. Here, we take the next step towards developing our characteristics-based filter design method, as described in section \ref{s:tunabilityApproach} and Fig. \ref{fig:approachschematic}. Specifically, in this section, we (a) define each of the filter characteristics as a function of the magnitude or phase of $\Pbold$ and then (b) use the expressions for magnitude and phase of $\Pbold_{sharp}$ from the previous section in order to derive expressions $\Psi = f(\Theta)$ for chosen filter characteristics in terms of the filter constants.

In what follows, we define the following set of characteristics, $\Psi$, in the normalized frequency domain: the peak frequency ($\Bcenter$), the group delay at the peak frequency ($N_\B$), any n-dB bandwidth ($\text{BW}_{n,\B}$) and associated quality factor ($Q_n$), the equivalent rectangular bandwidth (ERB) and associated quality factor ($Q_{erb})$, the convexity of the magnitude of the frequency response at the peak ($S_\B)$, and phase accumulation ($\phiaccum$). While our focus in what follows are the definitions and expressions as defined in the normalized frequency ($\B$), we also provide expressions for the characteristics as defined in regular frequency in the underbraces of the equations.

The peak (normalized) frequency is defined as,

\small
\begin{equation}
\begin{split}
    \Bcenter & \triangleq \underset{\B}{\mathrm{argmax}}  \, \level{\Pbold(\B)} \overset{\mathrm{sharp}}{=} \bp\\
    & \equiv \frac{1}{\text{CF}(x)} \underbrace{\underset{f}{\mathrm{argmax}}  \, \level{\Pbold(\B(f,x))} }_{\triangleq \fpeak} \;.
\end{split}
\label{eq:eqBpeak}
\end{equation}
\normalsize

which results in $\fpeak = \text{CF}(x)$ and $\Bcenter = 1$ which is the case by construction. Note that we may alternatively have chosen to define a central $\B$ based on phase slope criteria instead of magnitude criteria. If we use the sharp-filter approximate expressions, the two definitions for a central $\B$ - that based on peak magnitude and that based on maximum group delay, result in equivalent expressions.

The maximum group delay, $N_\B$, occurs at the peak frequency and is defined - in cycles, as,

\small
\begin{equation}
\begin{split}
    N_\B & \triangleq \frac{1}{2\pi} \max (-\frac{d\phase{\Pbold}}{d\B} ) \overset{\mathrm{sharp}}{=} \frac{1}{2\pi} \frac{\Bu}{\Ap}\\
   & \equiv \CFx \underbrace{\frac{1}{2\pi} \max (-\frac{d\phase{\Pbold}}{df} )}_{\triangleq N_f}\;.
\end{split}
\label{eq:eqNb}
\end{equation}
\normalsize

We note that $N_\B$ is the group delay at the peak frequency rather than the group delay at various frequency points as used by some other methods for filter design. 
However, in the case of GEFs (which are effectively linear-phase), group delay at the peak, maximum group delay, and group delay at frequencies with non-negligible magnitudes are all equivalent - as may be seen in Fig. \ref{fig:manySharpVsPhysP}. Consequently, one may simply only specify $N_\B$ rather than an entire set of group delays without concerns regarding signal distortion.

Phase accumulation in cycles (not radians) is equivalently defined in both $\B$ and $f$ domains and is \footnote{where max and min are over $\B \geq 0$},

\small
 \begin{equation}
     \phiaccum \triangleq -\frac{\max(\phase{\Pbold}) - \min(\phase{\Pbold})}{2\pi} \overset{\mathrm{sharp}}{=} \frac{\Bu}{2}\;.
     \label{eq:eqPhaseAccum}
 \end{equation}
\normalsize

It is the only non-peak-centric measure we take into consideration.


The n-dB bandwidth is,

\small
\begin{equation}
\begin{aligned}
    \text{BW}_{n,\B} & \triangleq \overbrace{\underset{\B > \Bcenter}{\arg } \{ \level{\Pbold}(\Bcenter) - \level{\Pbold}(\B) = n \} }^{\triangleq y_n^+} \\ & - \overbrace{\underset{\B < \Bcenter}{\arg } \{ \level{\Pbold}(\Bcenter) - \level{\Pbold}(\B) = n \} }^{\triangleq y_n^-} 
    \\& \overset{\textrm{sharp}}{=} 2\Delta_n =  2\Ap \sqrt{10^{\frac{n}{10\Bu}} - 1}
    \\
   & \equiv \frac{1}{\CFx} \text{BW}_{n,f}\;,
\end{aligned}
\label{eq:eqBWBn}
\end{equation}
\normalsize

where we obtained $\Delta_n$ by solving $n = \frac{20}{\log(10)} \big(-\frac{\Bu}{2} \log(\Ap^2) + \frac{\Bu}{2} \log(\Ap^2 + \Delta_n^2) \big)$.

The associated quality factor is,

\small
\begin{equation}
\begin{split}
    Q_n & \triangleq \frac{\Bcenter}{\BWndBbeta} \overset{\mathrm{sharp}}{=} \frac{\bp}{2\Ap} \big(10^{\frac{n}{10\Bu}} - 1\big)^{-\frac{1}{2}}\\
    & \equiv \frac{\fpeak}{\BWndBf} \;,
\end{split}
\end{equation}
\normalsize

which is equivalent regardless of whether it is defined in the $\B$ or $f$ domains.


The equivalent rectangular bandwidth (ERB) is,

\small
\begin{equation}
\begin{split}
    \text{ERB}_\B^{(\B_1, \B_2)} & \triangleq \frac{1}{{|\Pbold(\Bcenter)|^2}} \defint{-\B_1}{\B_2}{|\Pbold(\B)|^2}{\B} \\ & \quad \overset{\mathrm{sharp}, (-\infty, \infty)}{=\joinrel=} \quad \sqrt{\pi} \Ap \frac{\Gamma(\Bu - \frac{1}{2})}{\Gamma(\Bu)}\\
    & \equiv \frac{1}{\CFx} \underbrace{\frac{1}{{|\Pbold(\B(\fpeak, x))|^2}} \defint{-f_1}{f_2}{|\Pbold\B(f, x))|^2}{f}}_{\triangleq\text{ERB}^{(f_1, f_2)}_f} \;. \\
\end{split}
\label{eq:ERBeq}
\end{equation}
\normalsize

The limits of integration are relevant when numerically estimating the ERB from data over a finite domain. The ERB we derive from analytic expressions is $\text{ERB}_\B^{(-\infty, \infty)}$ and it is this version of the ERB we impose when we provide specifications to design filters. In contrast, the ERB over a finite range is the ERB computed from a designed filter.

To derive our expression for ERB, we first evaluate the indefinite integral which results in Gauss Hypergeometric Functions (GHF), $_2F_1(a,b;c;z)$ with $ z = - (\frac{\B-\bp}{\Ap})^2 $. We then express this in terms of GHFs in $z^{-1}$, and then expand those GHFs around zero to determine the expressions at $\B = \infty, -\infty$.  \removeInShortVer{Details are included in appendix \ref{s:appendixERBderivation}.}

The associated quality factor is,
\small
\begin{equation}
\begin{split}
    Q_{erb} & \triangleq \frac{\Bcenter}{\text{ERB}_{\B}} \overset{sharp}{=} \frac{\bp}{\sqrt{\pi} \Ap} \frac{\Gamma(\Bu)}{\Gamma(\Bu-\frac{1}{2})} \approx  \frac{\expn{b}\bp \Bu^{1-a}}{2\pi\Ap} \\
    & \equiv \frac{\fpeak}{\text{ERB}_{f}}
\end{split}
\label{eq:QerbEq}
\end{equation}
\normalsize

which is equivalent in both $\B$ and $f$ domains. The approximation is found empirically and is valid for relevant $\Bu$ values, $\Bu \geq \frac{3}{2}$, with fixed values for $b = 1.02, a = 0.418$.  





We introduce an additional magnitude-based filter characteristic, $S_{\B}$, that may be considered a more peak-centric measure of bandwidth. We define $S_\B$ in decibels as:

\small
\begin{equation}
\begin{split}
        S_\B & \triangleq - \frac{d^2\level{\Pbold(\B)}}{d\B^2} \Bigg|_{\Bcenter} \overset{sharp}{=} \frac{20}{\log(10)} \frac{\Bu}{\Ap^2} \\
        & \equiv \underbrace{-\frac{d^2\level{\Pbold(\B(\fpeak, x))}}{df^2}  \Bigg|_{\fpeak}}_{\triangleq S_f} \CFx^2 \;.
\end{split}
\label{eq:SB}
\end{equation}
\normalsize

As $\Ap$ decreases, the filters become sharper resulting in an increased $S_\B$, and equivalently, decreased  $\textrm{BW}_{n,\B}$ and ERB \footnote{We note that $S_\B$ is very useful for \textit{specifying} desired peak-centric filter behavior and designing filters based on this specification; however it is not appropriate to \textit{compute} this characteristic from any \textit{measured} data of filter transfer functions (at least not without smoothing) due to noise issues that are compounded by taking derivatives in \removeInShortVer{equation }(\ref{eq:SB}).}.

In this section, we have defined the frequency-domain filter characteristics of interest, $\Psi$, and derived their expressions in terms of the filter constants $\Theta$ using the sharp-filter expressions for magnitude and phase of GEFs. These expressions, $\Psi = f(\Theta)$, are summarized in Table \ref{tab:charInTermsOfParams}. Note that our definitions for $\phiaccum, Q_{n}, Q_{erb}$ are the same in $f$ and $\B$ domains. The other characteristics, $ \Bcenter, \textrm{BW}_{n,\B}, \textrm{ERB}_\B, N_\B, S_\B$ are easily converted into their counterparts in the frequency domains as per their above definitions. In what follows, we use $N$ in lieu of $N_\B$ for simplicity unless otherwise noted. 

\begin{table}
\caption{Filter characteristics as a function of filter constants}
    \label{tab:charInTermsOfParams}
\begin{center}
\setlength{\tabcolsep}{3pt}
        \begin{tabular}{|l|l|l|}
    \hline
     & \textbf{Filter Characteristics} & \textbf{Expression in Terms of Filter Constants} \\
    \hline
    I.1 & $\Bcenter = \frac{1}{\CFx} f_{peak}$ & $\bp$ \\
    \hline
    I.2 & $N_\B = \CFx N_f$ & $\frac{1}{2\pi} \frac{\Bu}{\Ap}$ \\
    \hline
    I.3 & $\phiaccum$ & $\frac{\Bu}{2}$ \\
    \hline
    I.4 & $Q_n$ & $\frac{\bp}{2\Ap} \big(10^{\frac{n}{10\Bu}} - 1\big)^{-\frac{1}{2}}$ \\
    \hline
    I.5 & $Q_{erb}$ & $\frac{\bp}{\sqrt{\pi} \Ap} \frac{\Gamma(\Bu)}{\Gamma(\Bu-\frac{1}{2})} \overset{\textrm{for } \Bu \geq \frac{3}{2}}{\approx} \frac{\expn{b}\bp \Bu^{1-a}}{2\pi\Ap}$ \\
    \hline
    I.6 & $S_\B = \textrm{CF}^2(x) S_f$ & $\frac{20}{\log(10)} \frac{\Bu}{\Ap^2}$ \\
    \hline

    \multicolumn{3}{p{251pt}}{Filter characteristics in terms of filter constants:  The expressions are derived from the sharp-filter approximations of the filter transfer function magnitude and phase. The filter characteristics without a $\B$ or $f$ subscript are equal in both domains. The constants for $Q_{erb}$ take on the following values: $b = 1.02, a = 0.418$.}\\
    \end{tabular}
    \end{center}    
\end{table}



We summarize the above expressions for filter characteristics in terms of filter constants in Table. \ref{tab:charInTermsOfParams}. Fig. \ref{fig:manySharpVsPhysP} shows that the dependence of $\Pbold$ magnitude and phase on constants $\Ap, \Bu$ is consistent with these expressions. Specifically, the figure shows that one may tune the behavior of GEFs ($\Pbold$) as follows: phase accumulation increases with $\Bu$, the maximum group delay increases with $\Bu$ and decreases with $\Ap$, sharpness of tuning decreases with $\Ap$ and increases with $\Bu$. 

\subsection{Parametizing the Filter in Terms of Filter Characteristics}
\label{s:parameterization}

We previously derived filter characteristics in terms of the filter constants, $\Psi = f(\Theta)$. As per section \ref{s:tunabilityApproach} and Fig. \ref{fig:approachschematic}, the next step towards parameterizing the GEF transfer functions in terms of filter characteristics, $\Pbold(s; \Psi)$, is to invert sets of the expressions $\Psi = f(\Theta)$ to derive expressions $\Theta = g(\Psi)$. In other words, we derive expressions for constants $\Theta$, as functions of subsets of $\Psi$ - including mixed magnitude-based and phase-based characteristics. These inversions are possible due to the simplicity of expressions in Table. \ref{tab:charInTermsOfParams} which we arrive at by using the sharp-filter approximation. We summarize the resultant expressions, $\Psi = g(\Theta)$, in Table. \ref{tab:char2param}. One particularly important combination (refer to row \ref{tab:char2param}.2 in Table. \ref{tab:char2param}) is in terms of the set of characteristics: $\Bcenter, N, Q_{erb}$. This set allows one to design sharp filters with large quality factors while introducing limited delay. The ability to simultaneous control quality factors and delays is due to the fact that the equations for $N$ and $Q_{erb}$ (or $Q_n$) in terms of $\Ap$ and $\Bu$ are independent of one another.

\begin{table*}[t]
\caption{Filter constants as functions of filter characteristics}
\label{tab:char2param}
\normalsize
\begin{center}
    \begin{tabular}{|l|l|l|l|p{0.25\linewidth}|}
    \hline
    & \backslashbox{$\mathbf{\Psi}$ \textbf{subset}}{ $\mathbf{\Theta}$ }   &  $\mathbf{\bp}$ &  $\mathbf{\Ap}$ &  $\mathbf{\Bu}$ \\ 
    \hline
    II.1& $\Bcenter, N, \phiaccum$    &  $\Bcenter$ &  $\frac{\phiaccum}{\pi N}$ &  $2\phiaccum$ \\ 
    \hline
    II.2&$\Bcenter, N, Q_{erb}$    &  $\Bcenter$ &  $ =\frac{1}{2\pi N} \Bu \quad$ or, \newline $= \frac{\Bcenter}{\sqrt{\pi} Q_{erb}} \frac{\Gamma(\Bu)}{\Gamma(\Bu-\frac{1}{2})}$ \footnote{The first option is more robust overall} &  $ \approx \expn{\frac{b}{a}}  \big(\frac{Q_{erb}}{\Bcenter N} \big)^{-\frac{1}{a}}$ \\ 
    \hline
    II.3&$\Bcenter, Q_{erb}, \phiaccum$    &  $\Bcenter$ &  $\frac{\Bcenter}{\sqrt{\pi} Q_{erb}} \frac{\Gamma(2\phiaccum)}{\Gamma(2\phiaccum -\frac{1}{2})}$ &  $2\phiaccum$ \\ 
    \hline
    II.4&$\Bcenter, Q_n, \phiaccum$    &  $\Bcenter$ &  $\frac{\Bcenter}{2Q_n} \big( 10^{\frac{n}{20\phiaccum}} - 1\big)^{-\frac{1}{2}}$ &  $2\phiaccum$ \\ 
    \hline
    II.5&$\Bcenter, S_\B, N$    &  $\Bcenter$ &  $\frac{40 \pi}{\log(10)} \frac{N}{S_\B}$ &  $\frac{80 \pi^2}{\log(10)} \frac{N^2}{S_\B}$ \\ 
    \hline
    II.6&$\Bcenter, S_\B, \phiaccum$    &  $\Bcenter$ &  $\sqrt{\frac{40 \pi}{\log(10)} \frac{\phiaccum}{S_\B}}$ &  $2\phiaccum$ \\ 
    \hline
    II.7&$\Bcenter, Q_n, N$    &  $\Bcenter$ &  $\frac{1}{2\pi N}\Bu$ & \footnotesize  solve the implicit equation $\frac{Q_n}{\Bcenter N} = \frac{\pi}{\Bu} \big( 10^{\frac{n}{10\Bu}} - 1\big)^{-\frac{1}{2}}$ 
    \\ 
    \hline
    \end{tabular}
     \end{center} 
     Filter constants in terms of sets of desired or measured filter characteristics: $\Gamma$ is the gamma function, and $a = 0.418, b = 1.02$. Note that our parameterizations allow for dictating both magnitude-based and phase-based filter characteristics. We have dropped the subscript, $\B$, from $N_\B$ for simplicity. \\
     $^{\mathrm{13}}$The first option is more robust overall
\end{table*}

In a simplification of Table. \ref{tab:char2param}, we may consider $\bp = 1$ as a fixed parameter - as it is used to set the peak frequency to be that prescribed, and just use appropriate combinations of two other characteristics for determining $\Ap$ and $\Bu$. We note that Table. \ref{tab:char2param} only includes a subset of possible combinations of filter characteristics that may be used to design GEFs. For instance, we may use the compound characteristics, $\frac{Q_{erb}}{Q_{10}}$ or $\frac{Q_{3}}{Q_{10}}$ (which are purely a function of $\Bu$) as part of our filter design method serving as an alternative to $\phiaccum$ or $\frac{Q}{\Bcenter N}$ for determining $\Bu$. This choice would allow for designing the filter based on only magnitude-based filter characteristics.


\subsection{Example Parameterizations}
\label{s:exampleparams}

As an example of parameterizing the GEFs in terms of filter characteristics, $\Pbold(s; \Psi)$, we consider the parameterization in terms of $\Bcenter, N,$ and $\phiaccum$ which results in the poles being $-\frac{\phiaccum}{\pi N} \pm i \Bcenter$, and the filter order being $-2\phiaccum$, or equivalently,

\small
\begin{equation}
    \Pbold(s) = \bigg(s^2 + 2\frac{\phiaccum}{\pi N}s + \Bcenter^2+ \big( \frac{\phiaccum}{\pi N} \big)^2\bigg)^{-2\phiaccum} \;.
    \label{eq:parametizationNandPhiaccum}
\end{equation}
\normalsize

Another example of characteristics-based parameterization of GEFs is using the set consisting of $\Bcenter, S_\B$, and $ N$,
\small
\begin{equation}
    \Pbold(s) = \bigg(s^2 + \frac{80 \pi N}{\log(10)S_\B} s + \Bcenter^2+ \big( \frac{40 \pi N}{\log(10)S_\B} \big)^2\bigg)^{-\frac{80 \pi^2 N^2}{\log(10)S_\B} } \;.
    \label{eq:parametizationNandS}
\end{equation}
\normalsize

One may very directly use such expressions for GEFs parameterized by filter characteristics (which we constructed using Table. \ref{tab:char2param}) for characteristics-based filter design.

\subsection{Verification: Analytic Testing against Expressions for Second Order System}
\label{s:validateDSHO}

Here we provide verification for the characteristics-based filter design methods and parameterizations for GEFs. We do so by verifying that the expressions we derived in section \ref{s:tunabilityCharsFxnOfParams} for filter characteristics in terms of filter constants, $\Psi = f(\Theta)$, for the particular case of $\Bu = 1$, are equivalent to those derived for the classically studied second order systems (which we review in appendix \ref{s:secondOrderToy}). To perform this verification, we consider:
\begin{enumerate}
    \item Our expressions for GEF characteristics in terms of constants - i.e. $\Psi = f(\Theta)$, as in Table. \ref{tab:char2param}, evaluated for $\Bu = 1$.
    \item Expressions from the classical literature for second order filter characteristics in terms of constants, which are derived for the case of lightly-damped oscillators.
\end{enumerate}

We compare the two aforementioned sets of expressions and determine whether or not they are equivalent. As previously mentioned, we are only interested in ranges of parameter values for which the system acts as a bandpass filter, and hence the set of filter characteristics derived in this paper (e.g. $Q_3$) can all be properly defined.

To compare whether or not the expressions for filter characteristics in terms of filter constants, $\Psi = g(\Theta)$, are equivalent for GEF with $\Bu = 1$ and for Classical Second Order Systems (CSOS), we must first maps the GEF constants ($\Ap, \bp$) to the CSOS constants ($\zeta$). Under the lightly damped condition, and comparing\removeInShortVer{equation } (\ref{eq:Pbold}) with \removeInShortVer{equation }(\ref{eq:soswwn}) we may express the GEF constants in terms of the CSOS constants as follows,

\small
\begin{equation}
    \begin{cases}
        \Ap = \zeta \;,\\
        \Ap^2 + \bp^2 = 1 \implies \bp = \sqrt{1 - \zeta^2} \;.
    \end{cases}
\end{equation}
\normalsize

Consequently, we may express the GEF characteristics for the case of $\Bu = 1$ (using Table. \ref{tab:charInTermsOfParams}) in terms of CSOS constants as,

\small
\begin{equation}
    \begin{cases}
        \phiaccum & = \frac{\Bu}{2} = \frac{1}{2} \\
        N_\B & = \frac{1}{2\pi} \frac{\Bu}{\Ap} = \frac{1}{2\pi} \frac{1}{\zeta} \\
        & \implies N_f = \frac{1}{\w_n \zeta} \\
        Q_3 & = \frac{\bp}{2\Ap} \bigg(10^{\frac{3}{10\Bu}} - 1\bigg)^{-\frac{1}{2}} \approx \frac{\sqrt{1-\zeta^2}}{2\zeta} \approx \frac{1}{2\zeta} \\
        \Bcenter & = \bp = \sqrt{1 - \zeta^2} \approx 1 \implies \w_r \approx \w_n \;.
    \end{cases}
\end{equation}
\normalsize

We find that above expressions for GEF characteristics for $\Bu = 1$ in terms of $\w_n$ and $\zeta$ are consistent with those in the classical literature for lightly-damped CSOS, thereby supporting the validity of our derivations.


\section{Accuracy of Filter Design Methods}
\label{s:evaluation}
In the previous section, we developed methods for the design of GEFs based on desired sets of values for filter characteristics which are the native and appropriate specification for this class of filters. As a result, it is inappropriate for us to assess the accuracy of our method in matching \textit{filter responses}. This is in contrast to various existing filter design methods for which the filter response of a particular length is the provided specification. Instead, we define the accuracy of the filter design method to be the degree to which the resulting filter fulfills the desired specifications on filter characteristics, $\Psi$. Here, we evaluate the accuracy of our filter design methods by numerically computing relative errors in filter characteristics over various ranges of values. Readers who are only interested in obtaining a sense of the high accuracy of the proposed methods - without the accompanying detailed analysis, may focus on the bar charts of Figs. \ref{fig:RespAndErrs6And005} and \ref{fig:RespAndErrs6And0052} that demonstrate the particularly low ranges of errors achieved in fulfilling the specifications when using the proposed filter design methods.



We numerically assess the accuracy of our filter design method by comparing the following two sets of filter characteristics, $\Psi$:

\begin{enumerate}
    \item $\Psi_{desired}$ which are specifications provided as values for sets of filter characteristics used to design $\Pbold(s)$. 
    \item  $\Psi_{num,P}$ which are filter characteristics numerically computed (according to the basic definitions in section \ref{s:tunabilityCharsFxnOfParams}) from the transfer function $\Pbold(s)$ after it has been designed using $\Psi_{desired}$.
\end{enumerate}

We define the relative error in each of the filter characteristics (abusing the vector notation) as,

\small
\begin{equation}
    \epsilon_{\Psi} = \frac{\Psi_{desired} - \Psi_{num,P}}{\Psi_{desired}} \;,
\end{equation}
\normalsize

and compute the relative errors over a range of values for each characteristic. We compute these errors and test the filter design methods for GEFs ($\Pbold$), then also compute these errors for one of the related bandpass filters, $\Vbold$.

\subsection{Relative Errors in Specifications of Designed GEFs}

\begin{figure}[tbp]
    \centering
    \includegraphics[width = \linewidth]{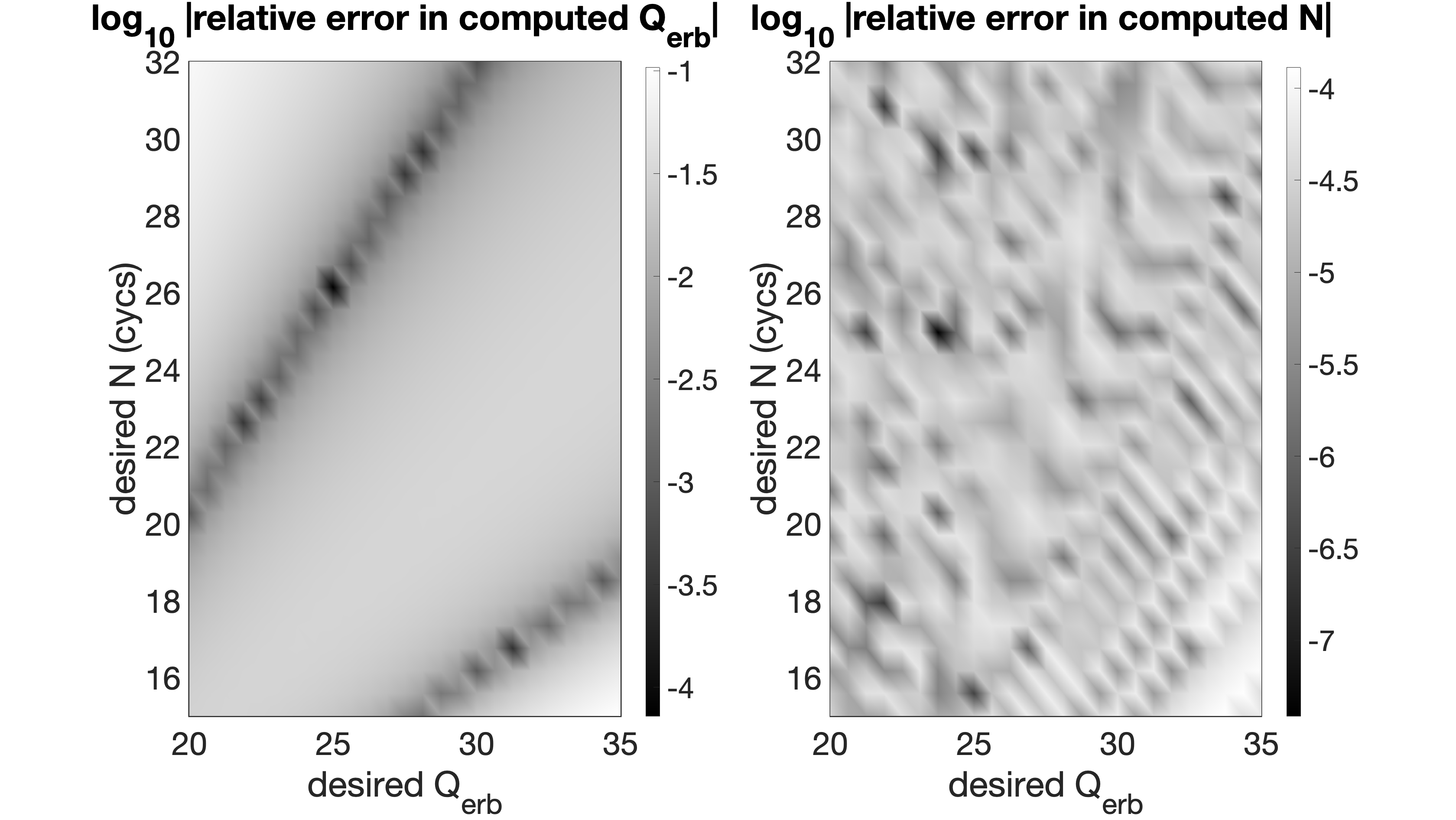}
    \caption{\textbf{The filter design method is highly accurate in fulfilling desired GEF characteristics}: The  error in computed characteristics for $Q_{erb}$ (left) and $N$ (right) are shown as a function of desired values for $Q_{erb}$ and $N$ that are most appropriate for AFs (we have imposed $\Bcenter = 1$ throughout). This figure is used to demonstrate the accuracy of the filter design method - as can be deduced from the range of values (in the colorbar), rather than to demonstrate any error pattern.}
    \label{fig:logErrorsVsDesiredQandN}
\end{figure}

We also provide the relative errors between desired and computed filter characteristics, $\epsilon_{\Psi}$, in the bottom panel of Figs. \ref{fig:RespAndErrs6And005} and \ref{fig:RespAndErrs6And0052}, (middle, red bars) for the various filter characteristics. We generated each of these figures using different sets of values for desired filter characteristics. Both figures show that the relative errors in characteristics are generally quite small (mostly $< 1.5 \%$) and hence that the design methods are highly accurate. Comparing Figs. \ref{fig:RespAndErrs6And005} and \ref{fig:RespAndErrs6And0052} illustrates that the relative errors in characteristics is smaller for sharper filters.

The relative errors in magnitude-based filter characteristics are small because these characteristics are peak-centric (e.g. BW$_{n,\B}, S_\B$) or weigh the peak region far more than it does the skirts (e.g. ERB). 
The relative error in $N_\B$ is particularly negligible as expected from phase-based characteristics which are nearly identical for $\Pbold$ and $\Pbold_{sharp}$. The relative error in $\phiaccum$ is larger than expected, but this is due to the fact that $\phi_{accum, num, P}$ is computed over a \textit{finite} range of $\B$. This becomes clear when we compute a relative error between desired filter characteristics and that computed from $\Pbold_{sharp}$ constructed from the desired characteristics: $\frac{\Psi_{desired} - \Psi_{num,Psharp}}{\Psi_{desired}}$ (notice the use of $P_{sharp}$ rather than $P$ in the subscript of the numerically computed characteristics). This relative error (left, blue bars in Figs. \ref{fig:RespAndErrs6And005} and \ref{fig:RespAndErrs6And0052}) is very nearly the same as that for the phase accumulation for $\Pbold$, thereby supporting our argument that the reason behind the error in phase accumulation is synthetic and due to finite ranges of $\B$ in computation rather than a true error.

\subsection{Relative Errors in Specifications of Designed $V$}

\begin{figure}[htbp]
    \centering
    \includegraphics[width = \linewidth]{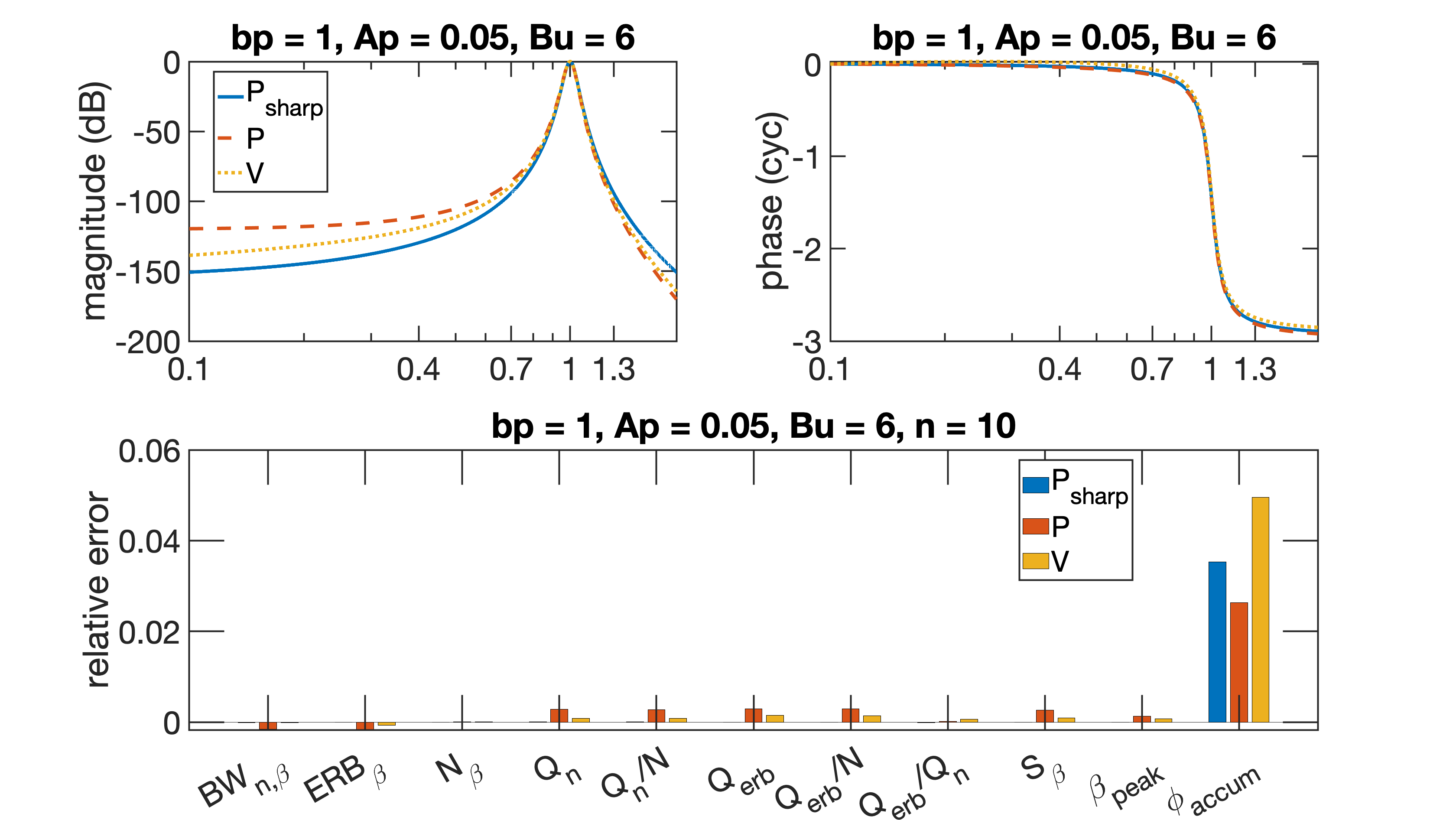}
    \caption{\textbf{The relative error in achieving desired filter characteristics is small}: The plot shows $\Pbold_{sharp}(\B), \Pbold(\B), \mathbf{V}(\B)$ designed using specified (desired) values for filter characteristics. 
    The magnitudes and phases (top) illustrate similarity near the peak, thereby supporting our choice of parametizing $\Pbold$ and $\mathbf{V}$ using peak-centric filter characteristics of $\Pbold_{sharp}$. 
    The bottom graph shows that the relative errors between the desired filter characteristics used to derive the filters and those computed from the generated transfer functions. This is done for $\Pbold_{sharp}$ (left, blue bars), $\Pbold$ (middle, red bars), and $\mathbf{V}$ (right, yellow bars). This figure was generated using filter constant values as per the figure title, or equivalently, using a subset of the following desired values for filter characteristics: $\Bcenter = 1, N_\B = 19.1 \textrm{ cyc}, \phiaccum = 3 \textrm{ cyc}, Q_{erb} = 25.9, \textrm{ERB}_\B = 0.039, Q_{10} = 14.6, \textrm{BW}_{10,\B} = 0.69, S_\B = 2.08 \times 10^4 \textrm{ dB}, \frac{Q_{erb}}{N} = 1.35 \textrm{ cyc}^{-1}, \frac{Q_{10}}{N} = 0.77  \textrm{ cyc}^{-1}, \frac{Q_{erb}}{Q_{10}} = 1.77$. An accompanying figure with an alternate set of filter characteristic values for parameterization is Fig. \ref{fig:RespAndErrs6And0052}.  }
    \label{fig:RespAndErrs6And005}
\end{figure}

\begin{figure}[htpb]
    \centering
    \includegraphics[width = \linewidth]{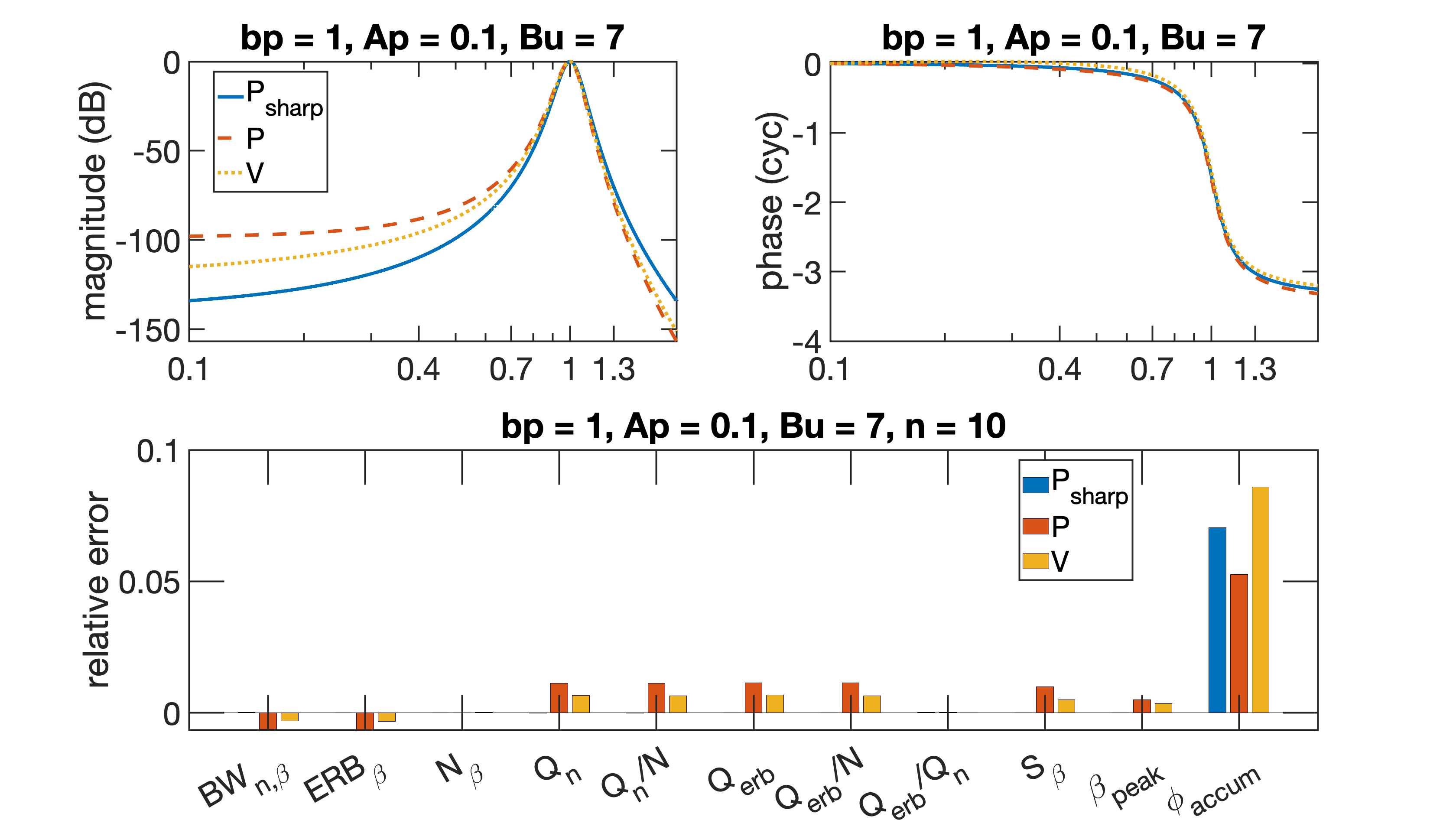}
    \caption{\textbf{The relative error in achieving desired filter characteristics is small}: See Fig. \ref{fig:RespAndErrs6And005} for a full description. This figure was generated using filter constant values (in the figure titles) corresponding to the following values of desired filter characteristic: $\Bcenter = 1, N_\B = 11.1 \textrm{ cyc}, \phiaccum = 3.5 \textrm{ cyc}, Q_{erb} = 14.1, \textrm{ERB}_\B = 0.071, Q_{10} = 8, \textrm{BW}_{10,\B} = 0.12, S_\B = 6.08 \times 10^3 \textrm{ dB}, \frac{Q_{erb}}{N} = 1.27 \textrm{ cyc}^{-1}, \frac{Q_{10}}{N} = 0.72  \textrm{ cyc}^{-1}, \frac{Q_{erb}}{Q_{10}} = 1.76$. Note that the filter is less sharp than that in Fig.  \ref{fig:RespAndErrs6And005} and so the relative errors are larger (though still very small).}
    \label{fig:RespAndErrs6And0052}
\end{figure}

We may also use the very same characteristics-based filter design methods that we developed in previous sections for GEFs to construct related filters such as $\Vbold$ (\ref{eq:V}). We do not need to separately derive a new filter-characteristics-based parameterization specifically for $\Vbold$. This is clearly the case for peak-centric characteristics-based parameterizations as seen in the similarity between $\Pbold$ and $\Vbold$ near the peak in the top panels of Figs. \ref{fig:RespAndErrs6And005} and \ref{fig:RespAndErrs6And0052}. We define relative errors between the specified $\Psi_{desired}$  and the $\Psi_{num,V}$ which we compute numerically from the generated transfer functions for $\Vbold$ using the definitions of section \ref{s:tunabilityCharsFxnOfParams}. We find that the relative errors in filter characteristics for $\Vbold$ (right, yellow bars in Figs. \ref{fig:RespAndErrs6And005} and \ref{fig:RespAndErrs6And0052}) are often even smaller than those for $\Pbold$ (middle, red bars) as the additional zero in $\Vbold$ decreases the magnitude asymmetry thereby making it closer to $\Pbold_{sharp}$.



\section{Extension to Multi-band Filters}
\label{s:multiband}

One method for designing dual-band and multi-band filters involves expressing these filters as the sum of single-band filters \cite{jankovic2015design}. Consequently, our filter design methods may be used for the design of multi-band filters that have distinct peaks provided that the constitutive single-band filters may be represented as sharp GEFs or related filters. 

\begin{figure}
        \centering
         \includegraphics[width = \linewidth]{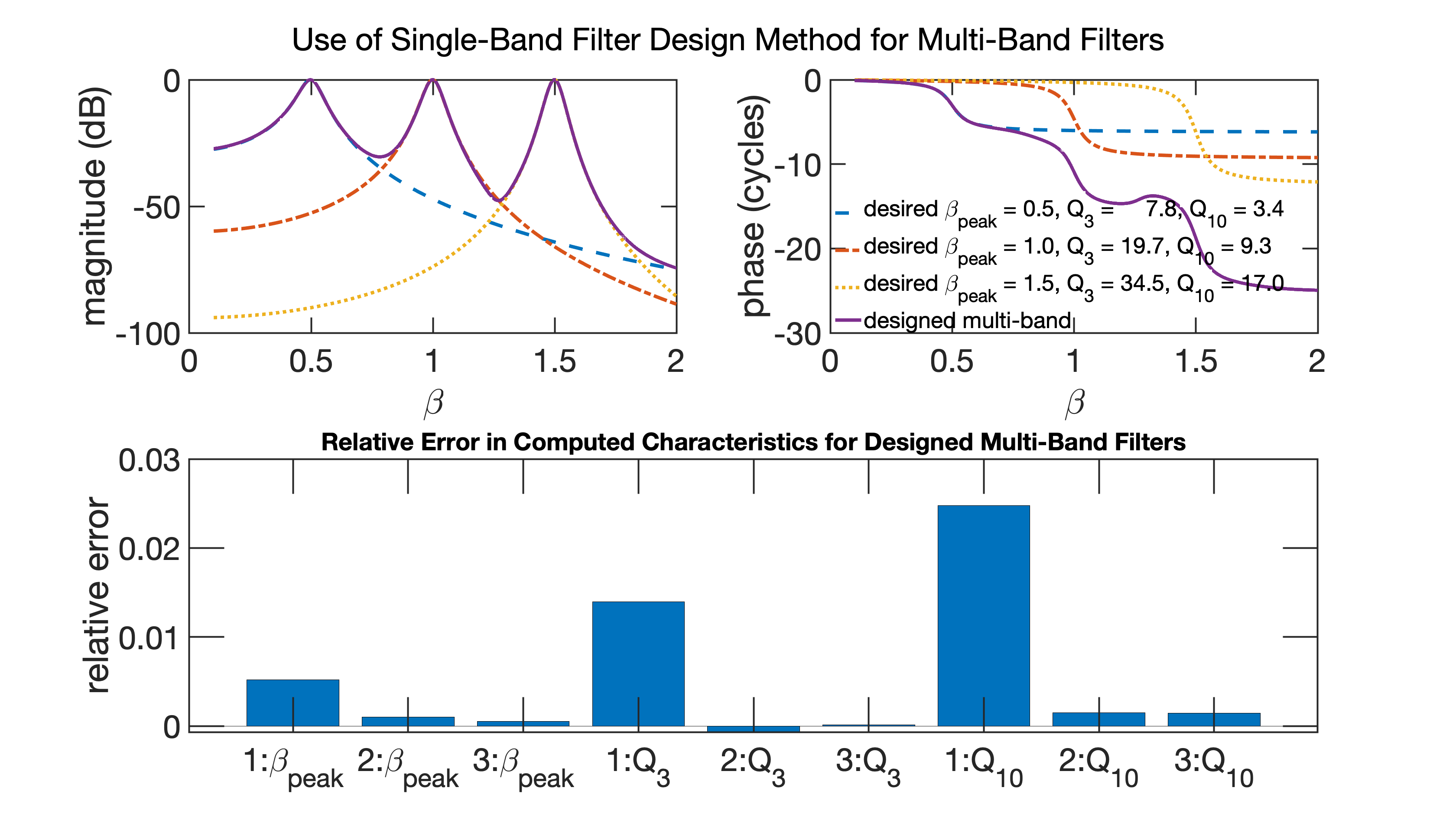}

    \caption{\textbf{The characteristics-based filter design method may be used for certain multi-band filters:} The multi-band filter (solid lines in top figure) is designed by summing the transfer functions of individual GEFs (dashed, dash-dotted, and dotted lines) which are in turn designed using specific values for the filter characteristics - $\Bcenter, Q_3, Q_{10}$. As in the legend, the desired characteristic values of the constitutive filters are: $\Bcenter = [0.5, 1.0, 1.5], Q_3 = [7.8, 19.7, 34.5], Q_{10} = [3.4, 9.3, 17.0]$. The relative error in the filter characteristics of the multi-band filter is shown in the bottom bar graphs.}
    \label{fig:multibandFig1}
\end{figure}

\begin{figure}
        \centering
         \includegraphics[width = \linewidth]{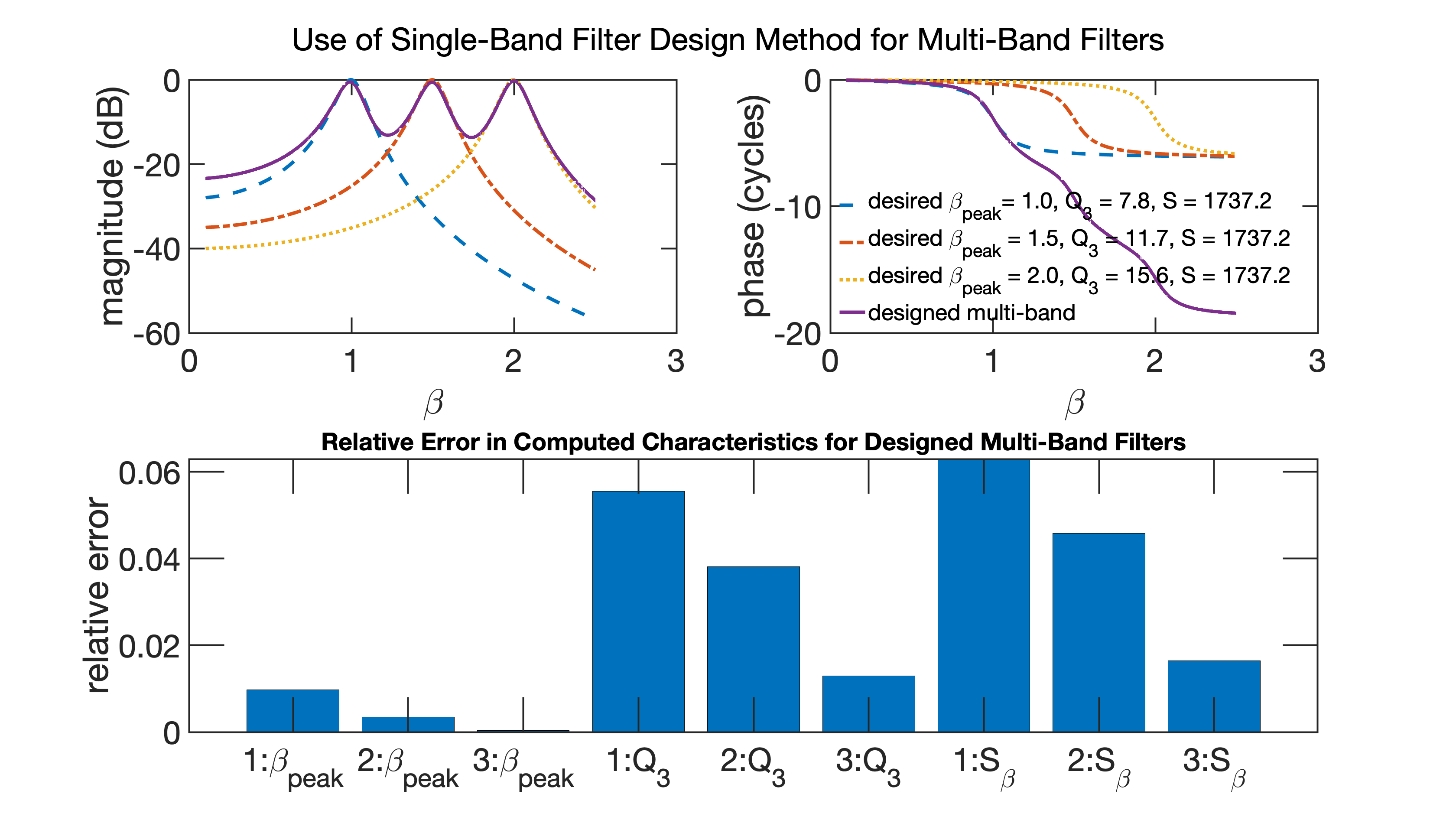}

    \caption{\textbf{The characteristics-based filter design method may be used for certain multi-band filters:} The multi-band filter (solid lines in top figure) is designed by summing the transfer functions of individual GEFs (dashed, dash-dotted, and dotted lines) which are in turn designed using specific values for the filter characteristics - $\Bcenter, Q_3,S_{\beta}$. As in the legend, the desired characteristic values of the constitutive filters are: $\Bcenter = [1, 1.5, 2], Q_3 = [7.8, 11.7, 15.6],S_{\beta} = [1737, 1737, 1737]$ . The relative error in the filter characteristics of the multi-band filter is shown in the bottom bar graphs.}
    \label{fig:multibandFig2}
\end{figure}

In Figs. \ref{fig:multibandFig1} and \ref{fig:multibandFig2}, we exemplify the frequency responses of multi-band filters constructed via the summation of constitutive GEFs that were designed using desired values for various sets of magnitude-based filter characteristics - specifically the set ($\Bcenter, Q_3, Q_{10}$) and the set ($\Bcenter, Q_3, S_\B$). For both multiband filters, we chose to set the peak magnitude of each of the constitutive filters to 0 dB. 

Using specifications on sets of magnitude-based filter characteristics to design multi-band filters is the appropriate choice for certain multi-band filters such as parametric equalizers, in which one may achieve finer control over the shapes of the response magnitudes. Alternative combinations may include characteristics that allow for control over frequency selectivities (e.g. BW, $Q, S_\B$) and delays ($N$). All sets of combinations naturally also specify peak frequencies and peak magnitudes.

We note that the choice of filter characteristics that may be used to design multi-band filters depends on the proximity of the peaks to one another. For instance, $Q_3$ may be better used in the design of multi-band filters compared with $Q_{10}$ particularly if the frequency responses of the constitutive single-band filters have more overlap. Of the magnitude-based selectivity characteristics,  $S_{\B}$, is the least restrictive in terms of requirements on peak-proximity and shape because it is the most peak-centric of these characteristics. Taking these observations into consideration when choosing the appropriate set of filter characteristics to design multi-band filters results in small relative errors as shown in Figs. \ref{fig:multibandFig1} and \ref{fig:multibandFig2}. Another added consideration that arises with multi-band filters is signal distortion which may occur if the peaks are close to one another and the group delays of constitutive filters are not chosen appropriately.

\section{Properties}
\label{s:features}
The proposed filter design methods achieve several of the desirable features described in section \ref{s:considerations} as demonstrated in the previous sections and summarized below. We group these properties under: natural specifications for GEFs, simultaneous specifications of various characteristics, accuracy and causal-stability, simplicity and computational efficiency, features for variable digital filter design and adaptive filtering applications, and finally, features for filterbank design.

\subsection{Natural Specifications for GEFs}

With our filter design methods, we are able to achieve many of the desired features for static filter design, variable filter design, and filterbanks outlined in section \ref{s:considerations}. The most important feature is that the methods construct filters directly based on the native set of specifications required by signal processing applications that use this class of filters - namely desired values for sets of filter characteristics. For these bandpass filters with a clear peak, supplying a frequency response over a length of frequencies rather than filter characteristics is quite an indirect approach. Furthermore, the expressions for parameterizations of GEFs with filter characteristics yield closed-form analytic expressions, thereby aiding in intuitive filter design. As previously mentioned, this was previously only achieved for first and second order filters.

\subsection{Simultaneous Specification of Various Characteristics}

Importantly, the parameterization allows for dictating magnitude and/or phase-based filter characteristics. The ability to simultaneously dictate the peak frequency, quality factor, and the group delay at the peak frequency (as shown in Table. \ref{tab:char2param}.2) is very powerful and importantly allows for designing sharply tuned filters without significant delay. The ability to simultaneously dictate peak frequency and any two types of bandwidths (as discussed in section \ref{s:parameterization}) allows for fine control over the shape of the magnitude response of the filter.

\subsection{Accuracy and Causal-Stability}

The proposed methods are highly accurate in achieving strict specifications on desired values of filter characteristics as discussed in section \ref{s:evaluation} and shown in Figs. \ref{fig:logErrorsVsDesiredQandN}-\ref{fig:multibandFig2}. The filter is inherently causally-stable by virtue of its parameterization by characteristics which are all positive. The positive values for characteristics necessarily lead to the pole pair being in the left half plane as can clearly be deduced by column $\Ap$ of Table. \ref{tab:char2param}.
In the discrete domain, the filter is also causally-stable provided that the appropriate analog to digital conversion techniques are used.

\subsection{Simplicity and Computational Efficiency}

The filter design methods are based on direct parameterizations of the filter in terms of characteristics - e.g. (\ref{eq:parametizationNandPhiaccum}) and \ref{eq:parametizationNandS}. Consequently, they are practically very simple to use, are maximally computationally efficient - due to the analytic closed-form expressions, and do not require iteration. The requirement of only needing to specify a set of filter characteristics rather than an entire frequency response over some length of frequencies also drastically reduces the number of required points needed in designing a filter to only three (in addition to the peak magnitude or gain constant).

\subsection{Features for Variable Digital Filter Design and Adaptive Filtering Applications}

The direct design methods allow for controlling the filter characteristics which becomes even more important for design problems in which the filter coefficients are adjustable. Consequently, our filter design approach is not only useful for static filter design but also stands as a particularly tractable form of variable filter design that may be used for applications such as adaptive filtering. Our parameterization is direct and simple (yet powerful) compared to existing methods for the design of variable digital filters which are categorized as frequency-transformation-based methods and multi-dimensional (in the spectral-parameter sense) polynomial approximation-based methods \cite{deng2004closed}. 

Our methods allow for varying all characteristics simultaneously rather than being limited to tuning only one or two characteristics as is the case with frequency-transformation-based methods for variable digital filters. We also avoid the potential issue of delay-free lines altogether. In contrast to multidimensional polynomial methods, our methods are computationally efficient, support a far broader range of values of filter characteristics, and can be used to dictate any of several sets of characteristics such as those in Table. \ref{tab:char2param} and those discussed in section \ref{s:parameterization}.


\subsection{Features for Filterbank Design}

The methods developed here are also useful for the design of uniform or nonuniform filterbanks due to features described previously. To apply our filter design methods to filterbanks, we must shift from $\beta$ to $f$ or $s$ to the Laplace $\sLaplace$. This is achieved by simply taking the $\Ap$ and $\bp$ of (\ref{eq:Pbold}) (that are determined from prescribed specifications on characteristics as in Table. \ref{tab:char2param}) and multiplying them by $f_{peak}$ \footnote{As done throughout this paper, we neglect the scale factor applied to the gain constant due to the transformation.}. The (regular) peak frequency, $f_{peak}$, must be prescribed for each filter in the filterbank. One way to prescribe $f_{peak}$ is provided in  (\ref{eq:BetaWithCFx}) which is the most suitable choice in the case of AFs. For the case of constant-Q filters, one may simply prescribe a single $Q$ for all filters in the filterbank. 

Features of the design methods that are useful for filterbanks include the simplicity and direct nature of the design methods, their computational efficiency, the ability to directly specify the peak frequencies and quality factors of constitutive filters and hence control cross-talk, and the ability to control both magnitude and phase-based characteristics simultaneously - thereby allowing one to achieve simultaneous requirements on frequency separation, selectivity, and synchronization or minimal delay. 


\section{Conclusions and Future Directions}
\label{s:conclusion}
\subsection{Contributions}

We have constructed methods to design IIR filters directly based on filter characteristics (which are the natural specifications for many applications and classes of filters). This eliminates the need to convert the specifications on characteristics to specifications on frequency responses. We develop these direct, non-iterative, characteristics-based design methods for the class of bandpass filters we refer to as GEFs and related bandpass and multi-band filters. Open source code for characteristics-based design of GEFs ($\Pbold$), $\Vbold$, and multiband filters may be found at \url{https://github.com/AnalyticModeling/GEFs}.

Our filter design methods satisfy several of the desired filter design features and considerations for bandpass filters with peaks - as discussed in sections \ref{s:considerations} and \ref{s:features}, in contrast to existing methods which generally satisfy some, but not all of these considerations together. The parameterization-based filter design methods are simple, computationally efficient, highly accurate, enable intuitive design, are inherently stable, and allow for higher-order behavior while retaining a small number of tunable characteristics. It allows for direct and fine control over characteristics, enables the construction of sharp filters with minimal delay, and is particularly accurate for sharp-filters.
The methods enable one to satisfy simultaneous specifications on a trio of magnitude and/or phase based characteristics. 


Future directions include (a) pursuing applications using filters appropriately designed using our methods, (b) developing filter design methods for GEFs and related filters that use alternative sets of filter characteristics, and (c) following the general paradigm introduced in this paper to develop characteristics-based methods for other classes of filters.

\subsection{Applications}

Proper filter design enables maximizing information obtained from processed signals. However, due to the absence of filter design methods that meet strict and multiple specifications on characteristics, the standard for many applications using GEFs (and related bandpass and multiband filters and filterbanks) has so far been to simply use a set of fixed filter constant values - e.g. fixed based on the Q-factor of the pole which is distinct from that of the filter, or based on a nominal search through values to get the desired filter behavior for a small subset of characteristics. The ability to parameterize the filters to directly control their behavior - as is the case with the characteristics-based filter design methods proposed here, allow us to better design, tune, and personalize filters for their intended applications such as those mentioned in section \ref{s:filterClass}.

Of these applications, the ones we consider most interesting include:  parametric equalizers constructed using our design methods for multiband filters, and signal classification problems that utilize well-designed filterbanks. We also expect that the parameterizations we derived for filter design may be useful (in filterbank capacities) for tunable multi-resolution modeling and signal processing. In the case of AFs, well-designed filters may also be used for the scientific study of the auditory system \cite{alkhairy2022cochlear}. The characteristics-based parameterized filters may further be considered beyond static filters from the perspective of higher-order bandpass variable digital filters and higher-order multiband variable digital filters. From this point of view, it is desirable to use our filter design methods for GEFs and related bandpass and multiband filters for adaptive filtering problems that benefit from optimizing over characteristics. This is motivated by the application of existing variable digital filters to adaptive filtering problems such as IIR notch adaptive filters \cite{lim2005piloted, cousseau2007factorized}, and generalized higher order variable-center frequency bandpass and bandstop filters \cite{koshita2013adaptive}. Using our methods, the adaptive algorithm would be over characteristics rather than over filter coefficients. 

\subsection{Alternative Sets of Filter Characteristics}

Some of the filter applications may benefit from design based on alternative sets of characteristics beyond our set of (predominantly) peak-centric frequency-domain characteristics. Consequently, future work may involve parameterizations using other relevant frequency-domain characteristics such as measures of the asymmetry of transfer function magnitudes. Additionally, it may be desirable to derive expressions for the filter constants in terms of time-domain impulse or step response filter characteristics such as settling time, instantaneous frequency, and parameters of envelope shape. These expressions would be useful for easily designing GEFs based on desired time-domain characteristics.

Taken together, the parameterization with time-domain characteristics and the parameterization with frequency-domain characteristics provide a mapping between characteristics in the time domain and those in the frequency domain. This is useful for building intuition and a common framework for interpreting observed characteristics in different domains. In addition, deriving mixed time and frequency-domain characteristics-based parameterizations would allow for designing filters with simultaneous requirements on time-domain and frequency-domain filter characteristics - e.g. simultaneous requirements on rise time and quality factor.

\subsection{Developing Characteristics-Based Design Methods for Other Filters}

Our filter design methods are developed for the class of GEFs but are also directly useful for related bandpass filters - e.g. $\Vbold$ and filters described by \removeInShortVer{equation }(\ref{eq:filterFormZero}), and associated multi-band filters. Beyond these classes of filters, the paradigm we followed to develop our analytic, deterministic, characteristics-based filter design methods  may provide a framework for guiding the development of characteristics-based filter design methods for other classes of filters for which the native specifications are also filter characteristics. Accordingly, future directions may involve developing methods for certain other bandpass filters used in signal processing and may further be developed for 2D bandpass filters such as those useful in image processing for which control of phase characteristics is very important. Additionally, it may be of interest to derive characteristics-based filter design methods for associated low-pass and high-pass filters as well as bandstop filters - especially notch filters for which fine control over the valley is desired.

\begin{appendices}
\section{Characteristics-Based Parameterization of Second Order Filters}
\label{s:secondOrderToy}

This appendix summarizes the commonly known characteristic-based parameterization of a class of second order bandpass filters. One particularly important, and unique, feature of bandpass filters describing Classical Second Order Systems (CSOS) is the ability to simply and directly design them based on desired frequency domain filter characteristics due to their parameterization by a natural frequency and quality factor. This is in contrast to iterative filter design methods that attempt to estimate filter coefficients based on a given desired frequency response and specifications.

The purpose of including this appendix is to (a) concretely illustrate characteristics-based parameterization and filter design using a well-known filter, (b) review the concept of normalized frequency, and (c) provide expressions that we use to verify the design methods we develop in the main text (section \ref{s:validateDSHO}). We consider the example of a lightly-damped driven harmonic oscillator (in its region of operation as a bandpass filter) which may be parameterized using the approximate 3-dB quality factor and natural frequency as,

\small
\begin{equation}
    H(\sLaplace) \propto \frac{1}{\sLaplace^2 + 2\zeta\w_n \sLaplace + \w_n^2} = \frac{1}{\sLaplace^2 + \frac{\w_n}{Q_3} \sLaplace + \w_n^2}  \;.
\end{equation}
\normalsize

This parametization enables designing filters for second order systems using a set of desired filter characteristics, $\Psi$ - which consists of: the half-power quality factor, $Q_3$, and the natural frequency, $\w_n$. The natural frequency is equivalent to the peak (resonant) frequency under the lightly damped oscillator condition $\zeta \ll 1$. We note that there are several conditions for various behaviors of CSOS. In order of least to most restrictive, these are: (a) $\zeta < 1$ which is the under-damped condition that results in a homogeneous solution oscillating at $\w_n \sqrt{1-\zeta^2}$; (b) $\zeta < \frac{1}{\sqrt{2}}$ which is the condition for significantly under-damped systems so that a resonance occurs for a forced system (at $\w_r = \w_n \sqrt{1-2\zeta^2}$) -- which still does not guarantee that it has a bandwidth and acts as a bandpass filter; (c) more restrictive conditions on $\zeta$ for the existence of bandwidths of various values of $n$; and (d) $\zeta \ll 1$ which is the condition for lightly damped oscillators which results in a bandpass filter with a peak / resonant frequency that may be approximated as the natural frequency, $\w_r \approx \w_n$.

An alternate representation of the CSOS where $s$ is defined based on normalized frequency, $\frac{\w}{\w_n}$ rather than on $\w$, only has a single degree of freedom, $Q_3$. Noting the shift from Laplace-$\sLaplace$ to normalized $s$ which is defined based on the frequency normalized to the natural frequency, $\w_n$, this becomes \footnote{We neglect the $\w_n$ scaling factor applied to the gain constant.},

\small
\begin{equation}
    H(s) \propto \frac{1}{s^2 + 2\zeta s + 1} = \frac{1}{s^2 + \frac{1}{Q_3} s + 1} \;.
    \label{eq:soswwn}
\end{equation}
\normalsize

Currently, such characteristics-based filter design is limited to second order bandpass filters and other particularly simple filters \footnote{Beyond first order low-pass and high-pass filters, we note efforts in the field of variable digital filters towards such characteristics-based parameterization of other filters after they are designed. The parameterizations are either limited to one or two characteristics and leave the rest of the filter coefficients free,  or involve computationally extensive estimation of multi-dimensional polynomial coefficients for which the polynomial is an approximation.}. However, as second order systems only have a single degree of freedom other than the natural frequency, their behavior is quite limited. For instance, the phase accumulation is fixed to half a cycle, the ratio of bandwidth to group delay is also fixed, as is the ratio of any two bandwidths of different levels (e.g. 3dB BW to 15dB BW). On the other hand, when using other filters with wider ranges of behavior to process signals, we lose the ability to design those filters based on desired filter characteristics.

The methods developed in the main text of this paper enable the characteristics-based design of classes of filters that have a wider range of behaviors than CSOS (specifically GEFs and related bandpass and multiband filters). The characteristics-based parameterizations for GEFs exemplified in  (\ref{eq:parametizationNandPhiaccum}) and (\ref{eq:parametizationNandS}) parallel the characteristics-based parameterizations for CSOS (\ref{eq:soswwn}).

\removeInShortVer{

\section{Sharp-filter Approximation and Conditions}
\label{s:appendixSharp}

In this section, we derive the sharp-filter approximation for the real and imaginary parts of $\kB$, which are in turn used to derive the sharp-filter expressions for the phase and magnitude of $\Pbold$ in section \ref{s:tunabilityOverview}, which are then in turn used for developing the filter design method that utilizes magnitude and/or phase-based filter characteristics. We also provide the condition for the validity of the sharp-filter approximation in this section.

\subsection{Derivation}
From equation, \ref{eq:wavenum}, we may express the real and imaginary parts of the $\kB$ separately as follows \footnote{In the corresponding mechanistic cochlear model $\kB = k \frac{l}{\B}$ denotes the dimensionless wavenumber which describes propagation in the transformed space $\B$, where $k$ is the complex wavenumber in $x$.}.



\small
\begin{equation}
\begin{aligned}
    \Im\{\kB\} & = \frac{1}{\Ap^2}\frac{- \Bu (\B - \bp)}{1 + (\frac{\B - \bp}{\Ap})^2} + \frac{1}{\Ap^2}\frac{- \Bu (\B + \bp)}{1+ (\frac{\B + \bp}{\Ap})^2}  \\ & \xrightarrow{\text{sharp-filter approx.}} \frac{- \Bu (\B - \bp)}{\Ap^2 + (\B - \bp)^2} \;, \\
    \Re\{\kB\} & = \frac{1}{\Ap^2}\frac{\Bu \Ap}{1 + (\frac{\B - \bp}{\Ap})^2} + \frac{1}{\Ap^2}\frac{\Bu \Ap}{1 + (\frac{\B + \bp}{\Ap})^2} \\ & 
    \xrightarrow{\text{sharp-filter approx.}} \frac{\Bu \Ap}{\Ap^2 + (\B - \bp)^2} \;.
\end{aligned}
\label{eq:rrealimag}
\end{equation}
\normalsize

Recall that $\B > 0$ and the parameters $\Bu, \bp, \Ap$ take on positive, real, values. For AFs, generally $\bp \approx 1, \Ap < 1$ and $\Bu \geq 2$ for integer-exponent GEFs. These separate expressions are then used to derive the expressions for the magnitude and phase of $P$ \footnote{as $P$ is defined as the exponent of the integral of the $\kB$} - see equations \ref{eq:Pmag} and \ref{eq:Pphase}, which are required towards developing the characteristics-based parameterization and filter design method. 

We may also express the complex $\kB$ as follows,

\begin{align}
    \kB = 2\Bu \frac{s+\Ap}{(s-p)(s-\pconj)} = \Bu \bigg( \frac{1}{s-p} + \frac{1}{s-\pconj} \bigg) \\ \xrightarrow{\text{sharp-filter approx.}} \Bu \frac{1}{s-p}  
    \;.
    \label{eq:wavenumSharp}
\end{align}

The above approximation leads to simple, interpretable counterparts to the full expressions, and enables us to invert filter characteristic expressions for parameters (which is needed for tunability and our filter design method). Note, however, that the approximations result in magnitudes for $\Pbold_{sharp}(\B)$ that are symmetric whereas the full expression, $\Pbold$, is asymmetric. Additionally, the sharp-filter approximation results in filters that are not physically realizable. Consequently, we use the approximation to develop tunability and parameterize the GEFs in terms of filter characteristics, but use the full expression to process signals and generate responses.


\subsection{Condition for Sharp-Filter Approximation}

The criteria for validity of the sharp-filter approximation depends on the values of $\B, \bp, \Ap, \Bu$ and can be determined from \removeInShortVer{equation }(\ref{eq:rrealimag}). In \removeInShortVer{equation }(\ref{eq:rrealimag}), $\frac{|\B - \bp|}{\Ap} < \frac{|\B + \bp|}{\Ap}$ is always true. Hence, the first term generally dominates. As this is especially true near the peak, the sharp-filter approximation is most accurate near the peak. Furthermore, the second term goes to zero much faster than the first if $\frac{|\B + \bp|}{\Ap} \gg 1$. Consequently, the sharp-filter approximation condition is $\Ap \ll \B + \bp$, which at the peak (when $\B = \bp$) is simply, $\Ap \ll 2\bp$. In other words, $\Ap \ll 2$ when $\bp = 1$. Practically, the sharp-filter approximation condition is satisfied by $\Ap < 0.2$. The  approximation is more appropriate for $\Re\{\kB\}$ and parts of variables derived from it (e.g. phase of $\Pbold$) than the imaginary part due to the nature of their respective numerators in \removeInShortVer{equation }(\ref{eq:rrealimag}). Consequently, the errors in achieving specified phase-based characteristics are generally smaller than those in achieving desired magnitude-based characteristics.

The condition on sharp-filter approximation may alternatively be determined from equations \ref{eq:wavenum} and \ref{eq:wavenumSharp}. From these expressions, it is clear that the sharp-filter approximation is valid when $|s-\pconj| \gg |s-p|$ (i.e. $\Ap$ is small or $\alpha \triangleq \frac{\Ap}{\beta + \bp} \ll 1$). Near the peak of $P$ (i.e. near $\B \approx \bp$), the sharp-filter condition is simply $\frac{\Ap}{2\bp} \ll 1$. This is consistent with our description in the previous paragraphs.

\removeInShortVer{\subsection{Behavior}

The sharp-filter expressions for $\Re\{\kB\}$ and $\Im\{\kB\}$ - \removeInShortVer{equation }(\ref{eq:rrealimag}), directly provide much information regarding the behavior of $P$ and its dependence on the parameters $\Ap, \bp, \Bu$:

\begin{itemize}
    \item $\bp$ determines the $\B$ at which the zero crossing of $\Im\{\kB\}$ and the maximum of $\Re\{\kB\}$ occur, which in turn determine the $\B$ at which both the peak in $|P|$ and maximum group delay occur
    \item $\Bu, \Ap$ determine the shape (widths and slopes) and scale of the magnitude and phase. This is better seen by noting that $\Bu, \Ap$ both determine the magnitude of $\Im\{\kB\}, \Re\{\kB\}$ at any particular $\B$ away from the peak (which corresponds to sharpness of $|P|$ and gradient of $\phi_P$ respectively)
    \item $\Bu$ determines the phase accumulation
    \item $\frac{\Bu}{\Ap}$ determines the maximum group delay (which is determined by the value of $\Re\{\kB\}$ at the peak of $|P|$)
\end{itemize}
}


\section{Derivation of ERB and $Q_{erb}$ Expressions}
\label{s:appendixERBderivation}
Here we derive our expressions for ERB (\removeInShortVer{equation }(\ref{eq:ERBeq}) and associated quality factor (\ref{eq:QerbEq}) in terms of the filter constants, $\Theta$.

The ERB, as defined in section \ref{s:tunabilityCharsFxnOfParams} may be expressed as,

\begin{equation}
\text{ERB}_{\infty}  = \frac{1}{|P_{max}|^2} I_\infty \;, 
\end{equation}

with,

\begin{equation}
    |P_{max}|^2  \overset{sharp}{\approx}\Ap^{-2\Bu}  \;,
\end{equation}

using the sharp-filter approximation for $|P(\bp)|^2$; and 

\small
\begin{equation}
    I_\infty  = \defint{-\infty}{\infty}{|P(\B)|^2}{\B} 
    I_\infty  = \lim_{n \xrightarrow{} \infty} J(\B) \bigg|_{-n}^{n} \;,
    \label{eq:IinfEq}
\end{equation}
\normalsize

with,

\small
\begin{equation}
    \begin{aligned}
    J(\B) & = \defint{}{\B}{|P(\B')|^2}{\B'} \\
    & \overset{sharp}{\approx} \defint{}{\B}{\bigg( (\Ap^2 + (\B'-\bp)^2 )^{-\Bu/2} \bigg)^2}{\B'} \quad \\
    & = \underbrace{  \frac{\B - \bp}{\Ap^{2\Bu}}  }_{\triangleq \zeta} \ _2F_1 (\frac{1}{2}, \Bu; \frac{3}{2}; - (\frac{\B-\bp}{\Ap})^2) \;,
\end{aligned}
\label{eq:J}
\end{equation}
\normalsize
using the sharp-filter approximation for $|P(\B)|^2$. This expression is in terms of the Gauss Hypergeometric Function (GHF), which - for the case of $|z|>1$ \footnote{which is the relevant form of the GHF in our case taking $\B \xrightarrow{} \infty, -\infty$}, is expressed as,

\small
\begin{multline}
    _2F_1(a,b;c;z)  = \frac{\Gamma(b-a) \Gamma(c)}{\Gamma(b) \Gamma(c-a)} (-z^{-1})^{a} \ _2F_1(a,a-c+1;a-b+1;z^{-1}) \\
    + \frac{\Gamma(a-b) \Gamma(c)}{\Gamma(a) \Gamma(c-b)} (-z^{-1})^{b} \ _2F_1(b, b-c+1; -a+b+1; z^{-1})   \;.
\label{eq:GHFlargez}
\end{multline}

\normalsize
To use \removeInShortVer{equation }(\ref{eq:GHFlargez}) for \removeInShortVer{equation }(\ref{eq:J}) of the ERB, we define $z$ and $y$ as
\begin{equation}
    z = -y = - (\frac{\B-\bp}{\Ap})^2 \;,
\end{equation}

with $|\B| \xrightarrow{} \infty \implies z \xrightarrow{} - \infty$, and with $a= \frac{1}{2} , b = \Bu, c= \frac{3}{2}$, and $\Gamma(1/2) = \sqrt{\pi}, \Gamma(3/2) = \sqrt{\pi}/2, \Gamma(1) = 1$.

As a result, we have,
\small
\begin{multline}
    _2F_1 (\frac{1}{2}, \Bu; \frac{3}{2}; -y) = \frac{\Gamma(\Bu-\frac{1}{2}) \frac{\sqrt{\pi}}{2} }{\Gamma(\Bu) 1} y^{-\frac{1}{2}} \ _2F_1(\frac{1}{2}, 0; \frac{3}{2} -\Bu; -y^{-1})\\
    + \frac{\Gamma(\frac{1}{2}-\Bu) \frac{\sqrt{\pi}}{2}}{\sqrt{\pi} \Gamma(\frac{3}{2}-\Bu)} y^{-\Bu} \ _2F_1(\Bu, \Bu - \frac{ 1}{2}; \Bu + \frac{1}{2}; -y^{-1})
\end{multline}
\normalsize

The first GHF in the above expression is simply $1$ due to the zero argument. We expand the second term using the formula for GHF for $|z|<1$,

\begin{equation}
    _2F_1(a,b;c;z) = 1 + \frac{ab}{c} z + O(z^2)
    \label{eq:GHFsmallzseries}
\end{equation}

and obtain,

\small
\begin{multline}
    _2F_1 (\frac{1}{2}, \Bu; \frac{3}{2}; -y) = a_1 y^{-\frac{1}{2}} \\ 
    + a_2 y^{-\Bu} \big( 1 + \Bu \frac{\Bu - \frac{1}{2}}{\Bu + \frac{1}{2}} (-y^{-1}) + O(y^{-2}) \big) \;,
\end{multline}
\normalsize

with,
\small
\begin{equation}
    \begin{aligned}
        a_1 & = \frac{\sqrt{\pi}}{2} \frac{\Gamma(\Bu-\frac{1}{2})  }{\Gamma(\Bu)}\\
        a_2 & = \frac{1}{2} \frac{\Gamma(\frac{1}{2} -\Bu ) }{ \Gamma(\frac{3}{2} -\Bu)} \;.
    \end{aligned}
\end{equation}
\normalsize

Substituting the above expansion for $_2F_1 (\frac{1}{2}, \Bu; \frac{3}{2}; -y)$ for large $y$ into \removeInShortVer{equation }(\ref{eq:J}) for $J(\B)$, we get,

\begin{multline}
    J(\B) = \frac{\B - \bp}{\Ap^{2\Bu}}  a_1 y^{-\frac{1}{2}} \\
    + a_2 y^{-\Bu} \big( 1 + \Bu \frac{\Bu - \frac{1}{2}}{\Bu + \frac{1}{2}} (-y^{-1}) + O(y^{-2}) \big) \;.
\end{multline}

As $|\B| \xrightarrow{} \infty$, because $\Bu > 1$, all but the first term go to zero, leaving,

\small
\begin{equation}
\begin{aligned}
    \lim_{|\B| \xrightarrow{} \infty} J(\B) & = \frac{\B - \bp}{\Ap^{2\Bu}}  a_1 y^{-\frac{1}{2}} \\
    & = \frac{\B - \bp}{\Ap^{2\Bu}}  a_1 ((\frac{\B - \bp}{\Ap})^2)^{-\frac{1}{2}} \\
    & =  \frac{\B - \bp}{\Ap^{2\Bu}}  a_1 \frac{\Ap}{|\B - \bp|} \\
    & = a_1 \Ap^{1-2\Bu} \frac{\B - \bp}{|\B - \bp|} \;.
\end{aligned}
\end{equation}
\normalsize

Therefore, 
\begin{equation}
         \lim_{\B \xrightarrow{} \infty} J(\B) = a_1 \Ap^{1-2\Bu} \;,
\end{equation}
and,
\begin{equation}     
         \lim_{\B \xrightarrow{} -\infty} J(\B) = -a_1 \Ap^{1-2\Bu} \;,
\end{equation}
which - by \removeInShortVer{equation }(\ref{eq:IinfEq}), result in,
\begin{equation}     
    I_\infty = 2a_1 \Ap^{1-2\Bu} = \sqrt{\pi} \frac{\Gamma (\Bu - \frac{1}{2})}{\Gamma(\Bu)}\Ap^{1-2\Bu} \,
\end{equation}
and consequently - as shown in \removeInShortVer{equation }(\ref{eq:ERBeq}),

\small
\begin{equation}   
    \text{ERB}_\infty = \sqrt{\pi} \frac{\Gamma (\Bu - \frac{1}{2})}{\Gamma(\Bu)}\Ap \;,
    \label{eq:ERBgamma}
\end{equation}
\normalsize

and its associated quality factor,
\begin{equation}     
         Q_{erb} = \frac{\bp}{\sqrt{\pi}\Ap} \frac{\Gamma (\Bu)}{\Gamma(\Bu - \frac{1}{2})} \;.
\end{equation}

While this representation is far superior to that using GHF as in \removeInShortVer{equation }(\ref{eq:J}), it may be further simplified to avoid using $\Gamma$ functions.

To do so, we note the following: let us consider the combination of filter characteristics, $\log( \mathrm{ERB} \times N$) (computed using equations \ref{eq:ERBgamma} and \ref{eq:eqNb}) which is purely a function of $\Bu$. We find that this compound characteristic is very well approximated by a line for the case $\Bu > \frac{3}{2}$ \footnote{It is important to note that we do not use the line approximation for the case of $\Bu = 1$ (second order systems). However, that is the single candidate integer point we are interested in below $\frac{3}{2}$. As a result, when expressing the filter characteristics in terms of the filter constants (to derive the inverse), we may use a piece-wise expression - with the value $ = \frac{\Gamma(\Bu)}{\Gamma(\Bu+ \frac{1}{2})} = \frac{2}{\sqrt{\pi}}$ for $\Bu = 1$ and the approximation derived above otherwise (for $\Bu \geq 2$). As a side note, we point out that the $n$ dB bandwidths derived in the main text do not require this different treatment of $\Bu = 1$.}. The best line for fitting to the above expression is,
\begin{equation}
    \log(\frac{Q_{erb}}{\Bcenter N}) = b-alog(\Bu), \quad \textrm{with } b = 1.02, a = 0.418 \;.
    \label{eq:QerbNlineFit}
\end{equation}

Substituting our expressions for $N$ - \removeInShortVer{equation }(\ref{eq:eqNb}) and $\Bcenter$ - \removeInShortVer{equation }(\ref{eq:eqBpeak}) in \removeInShortVer{equation }(\ref{eq:QerbNlineFit}), we arrive at,

\begin{equation}
    \log(Q_{erb}) = b + \log(\frac{\bp}{2\pi\Ap}) + (1-a)\log(\Bu),
\end{equation}

which we take the exponential of in order to arrive at our simplified expression for ERB and $Q_{erb}$ - \removeInShortVer{equation }(\ref{eq:QerbEq}) or table \ref{tab:char2param}).

}

\end{appendices}

\bibliography{references}

\begin{thebibliography}{10}

\bibitem{galvez2015time}
M.~F.~S. G{\'a}lvez, S.~J. Elliott, and J.~Cheer, ``Time domain optimization of filters used in a loudspeaker array for personal audio,'' {\em IEEE/ACM Transactions on Audio, Speech, and Language Processing}, vol.~23, no.~11, pp.~1869--1878, 2015.

\bibitem{oppenheim1999discrete}
A.~Oppenheim, R.~Schafer, and J.~Buck, {\em Discrete-time Signal Processing}.
\newblock Prentice Hall international editions, Prentice Hall, 2nd~ed., 1999.

\bibitem{cho2021design}
W.~Cho, D.~Chung, Y.~Kim, I.~Kim, and J.~Jeon, ``Design of fir half-band filter with controllable transition-band steepness,'' {\em IEEE Access}, vol.~9, pp.~52144--52154, 2021.

\bibitem{milic2022robust}
D.~N. Milic, E.~Avignon-Meseldzija, J.~A. Anastasov, H.~Meliani, and A.~Benlarbi-Dela{\"\i}, ``A robust algorithm for the design of wideband positive-slope linear group delay filters,'' {\em IEEE Transactions on Circuits and Systems I: Regular Papers}, vol.~69, no.~10, pp.~4258--4271, 2022.

\bibitem{pepe2022deep}
G.~Pepe, L.~Gabrielli, S.~Squartini, C.~Tripodi, and N.~Strozzi, ``Deep optimization of parametric iir filters for audio equalization,'' {\em IEEE/ACM Transactions on Audio, Speech, and Language Processing}, vol.~30, pp.~1136--1149, 2022.

\bibitem{nordebo2000use}
S.~Nordebo, I.~Claesson, and M.~Dahl, ``On the use of remes multiple exchange algorithm for linear-phase fir filters with general specifications,'' {\em IEEE Transactions on Education}, vol.~43, no.~4, pp.~460--463, 2000.

\bibitem{alkhairy1993design}
A.~S. Alkhairy, K.~G. Christian, and J.~S. Lim, ``Design and characterization of optimal fir filters with arbitrary phase,'' {\em IEEE Transactions on Signal Processing}, vol.~41, no.~2, pp.~559--572, 1993.

\bibitem{lu2017design}
W.-S. Lu and T.~Hinamoto, ``Design of least-squares and minimax composite filters,'' {\em IEEE Transactions on Circuits and Systems I: Regular Papers}, vol.~65, no.~3, pp.~982--991, 2017.

\bibitem{paperB2}
S.~A. Alkhairy, ``Rational-exponent filters with applications to generalized exponent filters,'' {\em IEEE Transactions on Circuits and Systems I: Regular Papers}, accepted.

\bibitem{jiang2020automatic}
T.~Jiang and J.~Zheng, ``Automatic phase picking from microseismic recordings using feature extraction and neural network,'' {\em IEEE Access}, vol.~8, pp.~58271--58278, 2020.

\bibitem{abdul2020hybrid}
Z.~K. Abdul, A.~K. Al-Talabani, and D.~O. Ramadan, ``A hybrid temporal feature for gear fault diagnosis using the long short term memory,'' {\em IEEE Sensors Journal}, vol.~20, no.~23, pp.~14444--14452, 2020.

\bibitem{karlos2020cochlea}
A.~Karlos and S.~J. Elliott, ``Cochlea-inspired design of an acoustic rainbow sensor with a smoothly varying frequency response,'' {\em Scientific Reports}, vol.~10, no.~1, p.~10803, 2020.

\bibitem{zhang2018underwater}
W.~Zhang, Y.~Wu, D.~Wang, Y.~Wang, Y.~Wang, and L.~Zhang, ``Underwater target feature extraction and classification based on gammatone filter and machine learning,'' in {\em 2018 international conference on wavelet analysis and pattern recognition (ICWAPR)}, pp.~42--47, IEEE, 2018.

\bibitem{harczos2012making}
T.~Harczos, A.~Chilian, and P.~Husar, ``Making use of auditory models for better mimicking of normal hearing processes with cochlear implants: the sam coding strategy,'' {\em IEEE transactions on biomedical circuits and systems}, vol.~7, no.~4, pp.~414--425, 2012.

\bibitem{sokolova2022real}
A.~Sokolova, D.~Sengupta, M.~Hunt, R.~Gupta, B.~Aksanli, F.~Harris, and H.~Garudadri, ``Real-time multirate multiband amplification for hearing aids,'' {\em IEEE Access}, vol.~10, pp.~54301--54312, 2022.

\bibitem{alias2016review}
F.~Al{\'\i}as, J.~C. Socor{\'o}, and X.~Sevillano, ``A review of physical and perceptual feature extraction techniques for speech, music and environmental sounds,'' {\em Applied Sciences}, vol.~6, no.~5, p.~143, 2016.

\bibitem{jankovic2015design}
N.~Jankovic, V.~Crnojevic-Bengin, P.~Meyer, and J.-S. Hong, ``Design methods of multi-band filters,'' {\em Advances in Multi-Band Microstrip Filters}, pp.~5--66, 2015.

\bibitem{feng2021filter}
W.~Feng, C.~Wang, X.~Chen, Y.~Shi, M.~Jiang, and D.~Wang, ``Filter realization of the time-domain average denoising method for a mechanical signal,'' {\em Shock and Vibration}, vol.~2021, pp.~1--13, 2021.

\bibitem{alkhairy2019analytic}
S.~A. Alkhairy and C.~A. Shera, ``An analytic physically motivated model of the mammalian cochlea,'' {\em The Journal of the Acoustical Society of America}, vol.~145, no.~1, pp.~45--60, 2019.

\bibitem{deepak2021convolutional}
B.~Deepak, S.~Verhulst, {\em et~al.}, ``A convolutional neural-network model of human cochlear mechanics and filter tuning for real-time applications,'' {\em Nature Machine Intelligence}, vol.~3, no.~2, pp.~134--143, 2021.

\bibitem{galbraith2008cochlea}
C.~J. Galbraith, R.~D. White, L.~Cheng, K.~Grosh, and G.~M. Rebeiz, ``Cochlea-based rf channelizing filters,'' {\em IEEE Transactions on Circuits and Systems I: Regular Papers}, vol.~55, no.~4, pp.~969--979, 2008.

\bibitem{koshita2017variable}
S.~Koshita, M.~Abe, and M.~Kawamata, ``Variable state-space digital filters using series approximations,'' {\em Digital Signal Processing}, vol.~60, pp.~338--349, 2017.

\bibitem{yu2011mixed}
Y.~J. Yu and W.~J. Xu, ``Mixed-radix fast filter bank approach for the design of variable digital filters with simultaneously tunable bandedge and fractional delay,'' {\em IEEE transactions on signal processing}, vol.~60, no.~1, pp.~100--111, 2011.

\bibitem{stoyanov1997variable}
G.~Stoyanov and M.~Kawamata, ``Variable digital filters,'' {\em J. Signal Processing}, vol.~1, no.~4, pp.~275--289, 1997.

\bibitem{nehorai1985minimal}
A.~Nehorai, ``A minimal parameter adaptive notch filter with constrained poles and zeros,'' {\em IEEE Transactions on Acoustics, Speech, and Signal Processing}, vol.~33, no.~4, pp.~983--996, 1985.

\bibitem{koshita2018recent}
S.~Koshita, M.~Abe, and M.~Kawamata, ``Recent advances in variable digital filters,'' {\em Digital Systems}, 2018.

\bibitem{yamaguchi2004adaptive}
K.~Yamaguchi, E.~Watanabe, and A.~Nishihara, ``Adaptive lowpass filters,'' in {\em The 2004 IEEE Asia-Pacific Conference on Circuits and Systems, 2004. Proceedings.}, vol.~1, pp.~509--512, IEEE, 2004.

\bibitem{pei2008fractional}
S.-C. Pei and H.-J. Hsu, ``Fractional bilinear transform for analog-to-digital conversion,'' {\em IEEE transactions on signal processing}, vol.~56, no.~5, pp.~2122--2127, 2008.

\bibitem{lyon1996all}
R.~F. Lyon, ``The all-pole gammatone filter and auditory models,'' in {\em Forum Acusticum'96, Antwerp}, 1996.

\bibitem{al2007novel}
M.~A. Al-Alaoui, ``Novel approach to analog-to-digital transforms,'' {\em IEEE Transactions on Circuits and Systems I: Regular Papers}, vol.~54, no.~2, pp.~338--350, 2007.

\bibitem{nelatury2007additional}
S.~R. Nelatury, ``Additional correction to the impulse invariance method for the design of iir digital filters,'' {\em Digital Signal Processing}, vol.~17, pp.~530--540, 2007.

\bibitem{goswami2021extended}
O.~P. Goswami, D.~K. Upadhyay, and T.~K. Rawat, ``Extended bilinear transform and multirate technique based approach for analog-to-digital transform,'' {\em International Journal of Electronics}, pp.~1--15, 2021.

\bibitem{paarmann2006mapping}
L.~D. Paarmann and Y.~H. Atris, ``Mapping from the s-domain to the z-domain via the phase-invariance method,'' {\em Signal processing}, vol.~86, no.~2, pp.~223--229, 2006.

\bibitem{van2003digital}
L.~Van~Immerseel and S.~Peeters, ``Digital implementation of linear gammatone filters: Comparison of design methods,'' {\em Acoustics Research Letters Online}, vol.~4, no.~3, pp.~59--64, 2003.

\bibitem{katsiamis2007practical}
A.~G. Katsiamis, E.~M. Drakakis, and R.~F. Lyon, ``Practical gammatone-like filters for auditory processing,'' {\em EURASIP Journal on Audio, Speech, and Music Processing}, vol.~2007, pp.~1--15, 2007.

\bibitem{alkhairy2017analytic}
S.~A. Alkhairy, {\em An analytic model of the cochlea and functional interpretations}.
\newblock PhD thesis, Massachusetts Institute of Technology, 2017.

\bibitem{alkhairy2022cochlear}
S.~A. Alkhairy, ``Cochlear wave propagation and dynamics in the human base and apex: Model-based estimates from noninvasive measurements,'' in {\em Nonlinearity and Hearing: Advances in theory and experiment: Proceedings of the 14th International Mechanics of Hearing Workshop}, AIP Publishing, 2022.

\bibitem{deng2004closed}
T.-B. Deng, ``Closed-form design and efficient implementation of variable digital filters with simultaneously tunable magnitude and fractional delay,'' {\em IEEE Transactions on Signal Processing}, vol.~52, no.~6, pp.~1668--1681, 2004.

\bibitem{lim2005piloted}
Y.~C. Lim, Y.~X. Zou, and N.~Zheng, ``A piloted adaptive notch filter,'' {\em IEEE Transactions on signal processing}, vol.~53, no.~4, pp.~1310--1323, 2005.

\bibitem{cousseau2007factorized}
J.~E. Cousseau, S.~Werner, and P.~D. Donate, ``Factorized all-pass based iir adaptive notch filters,'' {\em IEEE Transactions on Signal Processing}, vol.~55, no.~11, pp.~5225--5236, 2007.

\bibitem{koshita2013adaptive}
S.~Koshita, Y.~Kumamoto, M.~Abe, and M.~Kawamata, ``Adaptive iir band-pass/band-stop filtering using high-order transfer function and frequency transformation,'' {\em Interdisciplinary Information Sciences}, vol.~19, no.~2, pp.~163--172, 2013.

\end{thebibliography}


\end{document}